\documentstyle[12pt,epsf,epsfig,cite]{article}

\def\etalk{{ et al., }}
\newcommand{\scaption}[1]{\caption{\protect{\footnotesize  #1}}}

\def\r0{$\rho^{0}$}

\newcommand{\xpom}{x_{_{\rm I\!P}}}
\newcommand{\modt}{\mid\!t\!\mid}
\newcommand{\gapprox}{\stackrel{>}{_{\sim}}}
\newcommand{\lapprox}{\stackrel{<}{_{\sim}}}
\newcommand{\pom}{\rm I\!P}
\newcommand{\reg}{\rm I\!R}

\newcommand{\mx}{M_{_X}}
\newcommand{\my}{M_{_Y}}
\newcommand{\yda}{y_{_{\it DA}}}
\newcommand{\yjb}{y_{_{\it h}}}
\newcommand{\ysig}{y_{_{\Sigma}}}
\newcommand{\qssig}{Q^2_{_{\Sigma}}}
\newcommand{\alphapom}{\alpha_{_{\rm I\!P}}}
\newcommand{\alphareg}{\alpha_{_{\rm I\!R}}}

\newcommand{\ftpom}{F_2^{^{\rm I\!P}}(\beta,Q^2)}
\newcommand{\ftreg}{F_2^{^{\rm I\!R}}(\beta,Q^2)}

\newcommand{\fpomnt}{f_{_{{\rm I\!P}/p}}(\xpom)}
\newcommand{\fregnt}{f_{_{{\rm I\!R}/p}}(\xpom)}
\newcommand{\fint}{f^{I}(\xpom)}

\newcommand{\bpom}{B_{_{\rm I\!P}}}

\newcommand{\breg}{B_{_{\rm I\!R}}}
\newcommand{\etamax}{\eta^{\rm LAr}_{\rm max}}

\newcommand{\dthreesig}{\frac{{\rm d}^3 \sigma_{e p \rightarrow e X
    Y}}{{\rm d}x \,{\rm d}\beta\,{\rm d}Q^2}}
\newcommand{\dthreesigb}{\frac{{\rm d}^3 \sigma}{{\rm d}x \,
 {\rm d}\beta\,{\rm d}Q^2}}
\newcommand{\dfivesig}{\frac{{\rm d}^5 \sigma_{e p \rightarrow e X Y}}
{{\rm d}x\,{\rm d}\beta\,{\rm d}Q^2\,{\rm d}\my\,{\rm d}t}}
\newcommand{\rdfi}{R^{D(5)}(\beta,Q^2,\xpom,t,\my)}
\newcommand{\rdfib}{R^{D(5)}}
\newcommand{\ftdt}{F_2^{D(3)}(\xpom,\beta,Q^2)}
\newcommand{\ftdfi}{F_2^{D(5)}(\xpom,\beta,Q^2,\my,t)}
\newcommand{\ftdfib}{F_2^{D(5)}}
\newcommand{\ftdtb}{F_2^{D(3)}}
\newcommand{\ftdf}{F_2^{D(4)}(\xpom,\beta,Q^2,t)}
\newcommand{\gev} {\rm \,GeV}
\newcommand{\gevt} {\rm \,GeV^2}
\newcommand{\apres}{ \alphapom(0)=1.203\pm 0.020({\rm stat.})\pm 
 0.013({\rm sys.})  ^{+0.030}_{-0.035} ({\rm model})}
\newcommand{\amres}{\alphareg(0)=0.50 \pm 0.11 ({\rm stat.})\pm 0.11
({\rm sys.}) ^{+0.09}_{-0.10} ({\rm model})}
 1  1
 1

\def\Q2{$Q^2$}

\mathchardef\Lcur="324C

\def\qr2{{Q^2\over2}}
\def\q2{{Q^2}}

\def\be{\begin{equation}}
\def\ee{\end{equation}}
\begin{document}
\begin{titlepage}
\begin{flushleft}
{\tt DESY 97-158}\hfill {\tt ISSN 0418-9833} \\
{\tt August 1997}
\end{flushleft}
\vspace*{3.0cm}
\begin{center}\begin{LARGE}
  \boldmath {\bf Inclusive Measurement of Diffractive}\\ {\bf Deep-inelastic
    $ep$ Scattering} \\ \unboldmath
\vspace*{2.5cm}
H1 Collaboration \\
\vspace*{2.5cm}
\end{LARGE}

\end{center}

\begin{abstract}

  \noindent A measurement is made of the cross section for the process
  $ep \rightarrow eXY$ in deep-inelastic scattering with the H1
  detector at HERA.  The cross section is presented in terms of a
  differential structure function $\ftdt$ of the proton over the
  kinematic range $4.5 < Q^2 < 75\,{\rm GeV}^2$. The dependence of
  $\ftdtb$ on $\xpom$ is found to vary with $\beta$, demonstrating
  that a factorisation of $\ftdtb$ with a single diffractive flux
  independent of $\beta$ and $Q^2$ is not tenable.  An interpretation
  in which a leading diffractive exchange and a subleading reggeon
  contribute to $\ftdtb$ reproduces well the $\xpom$ dependence of
  $\ftdtb$ with values for the pomeron and subleading reggeon
  intercepts of $\alphapom(0)=1.203\pm 0.020({\rm stat.})\pm
  0.013({\rm sys.}) ^{+0.030}_{-0.035} ({\rm model})$ and
  $\alphareg(0)=0.50\pm 0.11({\rm stat.})\pm 0.11 ({\rm
    sys.})^{+0.09}_{-0.10} ({\rm model})$, respectively.  A fit is
  performed of the data using a QCD motivated model, in which parton
  distributions are assigned to the leading and subleading exchanges.
  In this model, the majority of the momentum of the pomeron must be
  carried by gluons in order for the data to be well described.
\end{abstract}
\vspace{2cm}
\begin{center}
{Submitted to Zeitshrift f\"{u}r Physik C} 
\end{center}
\end{titlepage}
\vfill
\clearpage
\begin{sloppypar}
\noindent C.~Adloff$^{35}$,                
 S.~Aid$^{13}$,                   
 M.~Anderson$^{23}$,              
 V.~Andreev$^{26}$,               
 B.~Andrieu$^{29}$,               
 V.~Arkadov$^{36}$,               
 C.~Arndt$^{11}$,                 
 I.~Ayyaz$^{30}$,                 
 A.~Babaev$^{25}$,                
 J.~B\"ahr$^{36}$
% ...
 \\
\bigskip 

\noindent
{\footnotesize{
% ...
$ ^{11}$ DESY, Hamburg, Germany$^a$ \\
% ...
$ ^{13}$ II. Institut f\"ur Experimentalphysik, Universit\"at Hamburg,
                                     Hamburg, Germany$^ a$  \\
% ...
$ ^{23}$ Physics Department, University of Manchester,
                                          Manchester, UK$^ b$\\
% ...
$ ^{25}$ Institute for Theoretical and Experimental Physics,
                                                 Moscow, Russia \\
 $ ^{26}$ Lebedev Physical Institute, Moscow, Russia$^ f$ \\
% ...
$ ^{29}$ LPNHE, Ecole Polytechnique, IN2P3-CNRS,
                             Palaiseau, France \\
 $ ^{30}$ LPNHE, Universit\'{e}s Paris VI and VII, IN2P3-CNRS,
                              Paris, France \\
% ...
$ ^{35}$ Fachbereich Physik, Bergische Universit\"at Gesamthochschule
               Wuppertal, Wuppertal, Germany$^ a$ \\
 $ ^{36}$ DESY, Institut f\"ur Hochenergiephysik,
                              Zeuthen, Germany$^ a$\\
% ...
$ ^b$ Supported by the UK Particle Physics and Astronomy Research
       Council, and formerly by the UK Science and Engineering Research
       Council \\
% ...
$ ^f$ Supported by the Deutsche Forschungsgemeinschaft \\
% ...
}}

\end{sloppypar}

\newpage
\section{Introduction}

For a complete understanding of the strong interaction it is necessary not
only to survey short range, hard partonic interactions, but also the long
range soft interactions that make up the bulk of hadron-hadron and
photon-hadron cross sections at high energy. The lack of a hard scale in such
soft processes makes calculations within quantum chromodynamic theory (QCD)
problematic, and so these processes are usually described using the
phenomenology of Regge theory~\cite{Regge} by the $t$-channel exchange of
mesons~\cite{Chew} and, at high energy, by the leading vacuum singularity,
the pomeron\footnote{In this paper the terms `pomeron exchange' and
  `diffraction' are used synonymously.} ($\pom$)~\cite{chew:pomeron}.
Measurements of deep-inelastic  diffraction, made possible for the
first time with the HERA electron-proton ($ep$)
collider\footnote{HERA is able to collide both
  positrons and electrons with protons. In this paper the term `electron' is
  used to describe generically electrons or positrons.}, offer
a unique chance to study the soft scattering process with a hard virtual
photon probe. It is hoped that precision measurements of  diffraction
in deep-inelastic scattering (DIS)
will guide the formulation of a `reggeon field theory' that
encompasses both hard and soft aspects of the strong 
force~\cite{lipatov,white} and
will therefore provide a more complete understanding of QCD.

Following the observation of deep-inelastic scattering events with a rapidity
gap~\cite{ref:zeus_lrg,ref:h1_lrg}, first measurements were made of the
contribution of such events to the DIS cross section, quantified in terms of
the differential structure function
$\ftdtb$~\cite{ref:h1_f2d_93,ref:zeus_f2d_93}. These measurements
demonstrated that the rapidity gap events could be attributed to an $ep$
process in which the interaction is dominantly diffractive. No substantial
$Q^2$ dependence of $\ftdtb$ was observed, thus indicating that the
substructure of diffractive exchange was of a point-like,
presumably partonic, nature. 

Presented here is a new measurement of $\ftdtb$, based upon the data
collected by H1 during 1994. The measurement is made using data in
which a forward rapidity gap is observed so that the hadronic final
state is partitioned into two distinct subsystems $X$ and $Y$, where
$Y$ is the system with rapidity closest to that of the incident
proton. System $X$ is measured with the main components of the H1
detector, whereas system $Y$ is not detected directly but the location
of the rapidity gap implies its mass is
less than $1.6\,{\rm GeV}$. 
The data amount to a factor $\sim10$ increase in statistics compared to our
previous analysis~\cite{ref:h1_f2d_93} and they permit a substantially more
accurate measurement of $\ftdtb$ over a wider kinematic range. Further, by
utilising data taken with the $ep$ interaction vertex shifted by
$\sim70\,{\rm cm}$ along the proton beam direction, the measurement is
extended to considerably lower values of $Q^2$ than hitherto possible.

\section{Cross Section Definition and Kinematic Variables}
\label{sec:form}         
\begin{figure}[ht]
   \begin{center}
\epsfig{figure=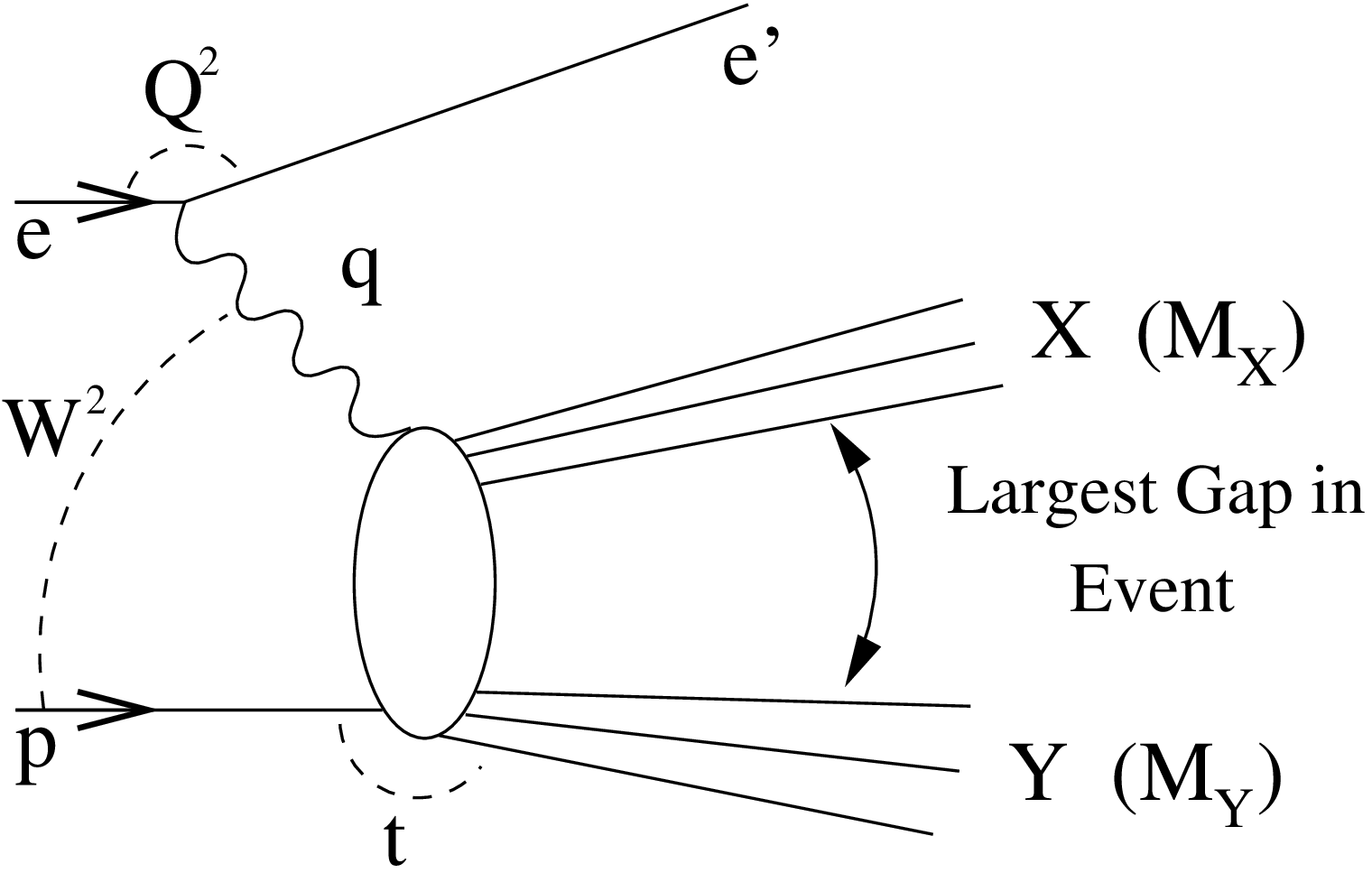,angle=270,width=0.8\textwidth}
   \end{center}
   \scaption{Schematic of an DIS scattering event $ep \rightarrow eXY$.}
\label{wsfig1}
\end{figure}

In the analysis presented in this paper the cross section is defined
in terms of a topological decomposition of the final state hadrons.
The hadronic final state, defined as all final state particles
excluding the scattered electron, is separated into two distinct
systems labelled $X$ and $Y$.  These are taken to be the two hadronic
systems which are separated by the largest interval in rapidity
between final state hadrons, where rapidity is calculated in the
virtual photon-proton centre of mass frame.  The system closer in
rapidity to that of the incident proton is taken to be the subsystem
$Y$. A schematic diagram of an $ep$ scattering, with this decompostion
of the final state, is shown in
figure~\ref{wsfig1}.

The kinematics of the process can be described by the invariant masses
of the two subsystems, $\mx$ and $\my$, and the Lorentz scalars
\begin{equation}
  x=\frac {-q^2}{2P\cdot q},\,\,\,\,\,\,\, 
  y=\frac {P\cdot q}{P\cdot k},\,\,\,\,\,\,\,  
 Q^2 = -q^2, \,\,\,\,\,\,\,  t=(P-p_Y)^2 .
           \label{eq:definition}
\end{equation}
Here $P$ and $k$ are the $4$-momenta of the incident proton and
electron respectively, $p_Y$ is the $4$-momentum of subsystem $Y$ and
$q$ is the $4$-momentum of the exchanged virtual photon ($\gamma^*$)
coupling to the electron.  The quantities $x$, $y$ and $Q^2$ are the
usual variables used to
describe DIS; $t$ is the square of the 4-momentum
transferred at the proton-$Y$ vertex.  Further useful variables are
\begin{equation}
  \beta = \frac{-q^2}{2q\cdot (P-p_Y)} =\frac{Q^2}{Q^2+\mx^2-t},
            \label{eq:betaQ2} 
\end{equation}
\begin{equation}
{\rm and}  \\ \,\,\,\,  \xpom =\frac{q\cdot (P-p_Y)}{q\cdot
    P}=\frac{Q^2+\mx^2-t}{Q^2+W^2-M_p^2} = \frac{x} {\beta}.
           \label{eq:xpomQ2}
\end{equation}
Here $W^2=(q+P)^2$ is the square of the centre of mass energy of the
virtual photon-proton system and $M_p$ is the proton mass.  

The decomposition of the final state has been chosen to most clearly
differentiate between various exchange mechanisms. Colourless
exchanges can produce states in which $X$ and $Y$ are clearly
separated in rapidity and both $\mx$ and $\my$ are small compared with
$W$. For colour exchange processes, however, the location of the gap
is defined by random fluctuations, the average gap size is small and
one or both of $\mx$ and $\my$ is generally large. In this paper
events are selected (see section~\ref{sec:eventsel}) such that
$\xpom<0.05$ and $\my<1.6\,{\rm GeV}$.  This ensures that both $\mx
\ll W$ and $\my \ll W$, so that $X$ and $Y$ are well separated and the
cross section is dominated by colourless exchange processes.  In the
infinite momentum frame of the incident proton, $\xpom$ may be
interpreted as the ratio of the momentum carried by the colourless
exchange to that of the incident proton, and $\beta$ as the ratio of
the momentum carried by the quark coupling to the virtual boson to
that of the colourless exchange.

In principle it is  possible to measure a differential cross section with
events decomposed in the scheme discussed above as a function of five
independent variables, and to define a differential structure function
$\ftdfi$ 
\begin{equation} 
\dfivesig =
\frac{4\pi\alpha_{\it em}^2}{\beta^2 Q^4}\left(1-y+
      \frac{y^2}{2(1+\rdfib)}\right)\ftdfib.
           \label{eq:F2D5}
\end{equation}
Here $\alpha_{\it em}$ is the fine structure constant and $\rdfi$ is
the ratio of the longitudinal to the transverse photon cross sections.
For the data used in this paper, however, the system $Y$ is not
measured directly, so that a differential measurement as a function of
$\my$ and $t$ is not possible.  Instead a measurement of the triple
differential cross section $\dthreesig$ is made, where an implicit
integration is performed over the two unmeasured quantities such that
$\my<1.6\,{\rm GeV}$ and $|t| <1\,{\rm GeV}^2$.  The cross section is
chosen to include, and is expected to be dominated by, the single
dissociative process $\gamma^*p \rightarrow X p$.  The differential
structure function $\ftdt$ is defined and, following the usual
convention~\cite{ref:DL_F2Pom,Ingprytz,ref:h1_f2d_93}, is extracted
with the assumtion of $\rdfib=0$.
\begin{equation} 
  \dthreesig = \frac{4\pi\alpha_{\it em}^2}{\beta^2
    Q^4}(1-y+\frac{y^2}{2})\ftdtb.
           \label{eq:F2D3}
\end{equation}
The influence  of a non-zero value of $R^{D(5)}$ is discussed further in
section~\ref{sec:phenfit}.

\section{ Simulation of Deep-inelastic {\boldmath $ep$} Interactions}
\label{sec:mc}
In this analysis Monte Carlo simulations are used to correct the data for the
effects of losses and smearing. A mixture of several models is used in order
to best reproduce the data since no single event generator gives a complete
description of all possible DIS final states.  Each generated event sample
includes a full simulation of the H1 detector and is passed through the same
analysis chain as the data.

The RAPGAP generator~\cite{rapgap} is used to generate the process
$ep\rightarrow eXp$ in the kinematic region $\mx> 1.1\,{\rm GeV}$ and
$\xpom<0.1$. The model has two components in which the proton couples
to a pomeron ($\pom$) or a meson ($\reg$), which subsequently
undergoes a hard collision with the incident electron.  The pomeron
and meson fluxes are taken to follow a simple Regge motivated form
with an exponential $t$ dependence so that
\begin{equation}
  \ftdf=\frac{e^{\bpom t}}{\xpom^{2\alphapom(t)-1}}\ftpom+\frac{e^{\breg
      t}}{\xpom^{2\alphareg(t)-1}}\ftreg,
\end{equation}
where the pomeron and meson trajectories are assumed to have the
linear form, $\alpha(t)=\alpha(0)+\alpha^{\prime}t$. The values for
the intercepts $\alphapom(0)=1.18$ and $\alphareg(0)=0.6$ are chosen
such that $\ftdt$ from the generator is in close agreement with the
results of the phenomenological fit presented in
section~\ref{sec:phenfit}. The values for the slope parameters,
$\bpom$ and $\breg$, and the shrinkage parameters,
$\alphapom^{\prime}$ and $\alphareg^{\prime}$, are not constrained by
this analysis. The value assumed for the slope parameter of the
pomeron, $\bpom=6\,{\rm GeV}^{-2}$, matches the value obtained in a
DIS measurement in which the leading proton was
detected~\cite{zeus_LPS_94}.  Since the shrinkage in the forward peak
of the diffractive cross section has not been measured at high $Q^2$
and, since the value used in the generator leads to negligible changes
to the measurements presented in this paper, $\alphapom^{\prime}$ is
set to zero.  The assumed values of $\breg=2\, {\rm
  GeV}^{-2}$ and $\alphareg^{\prime}=0.9\,{\rm GeV}^{-2}$ are taken
from hadron scattering experiments~\cite{kaidalov:review,apel:chexch}.

The pomeron structure function $\ftpom$ assumed in the generator is
derived from parton distributions taken from a QCD fit to the present
data similar to that described in section~\ref{sec:qcdfit}.  The meson
structure function $\ftreg$ is taken from the GRV parameterisation of
the structure function of the pion~\cite{GRVPion}.  The structure
functions are assumed to evolve with $Q^2$ according to the leading
order DGLAP evolution equations~\cite{DGLAP}.  In addition to the
{${\cal{O}}(\alpha_{em})$} quark-parton model diagram ($eq \rightarrow
eq$), boson-gluon fusion ($eg \rightarrow eq\bar{q}$) and QCD-Compton
($eq \rightarrow eqg$) processes are generated according to the
{$\cal{O}$}($\alpha_{em}\alpha_s$) matrix elements.  Higher order QCD
corrections are provided either by the colour dipole model as
implemented in ARIADNE~\cite{ariadne} or parton showers similar to
those used in simulations of inclusive DIS (for example see
LEPTO~\cite{lepto}). QED radiative processes are included via an
interface to the program HERACLES~\cite{heracles}.

The low mass region of diffractive production ($\mx < 1.1\,{\rm GeV}$)
is modelled with the DIFFVM~\cite{diffvm} generator, which simulates
the production of the vector mesons $\rho(770)$, $\omega(782)$ and
$\phi(1020)$. The cross sections are taken from measurements for
$\rho(770)$ and $\phi(1020)$ production in
DIS~\cite{pierre,nmcrho,zeusphi,h1phi} and a ratio of
$\omega(782):\rho(770)$ of $1:9$ as given by quark counting rules.
The $Q^2$ and $W$ dependences of the cross sections are assumed to
follow those measured for $\rho(770)$ production~\cite{pierre}.

The DIFFVM generator is also used to model proton dissociation.  The
cross section for proton dissociation is assumed to have a ${\rm d}
\sigma / {\rm d} \my \propto 1/\my^2$ dependence at large $\my$, with
a parameterisation based on measurements of proton dissociation on
deuterium~\cite{pdissdeuteron} for the low mass region. Two different
decay mechanisms are considered: decay with limited $p_T$ following
the Lund fragmentation scheme~\cite{lund}, or isotropic phase space
decay with multiplicity following KNO scaling laws~\cite{kno}. The
ratio of proton dissociation to elastic proton interactions is assumed
to be $1$.

The kinematic region of high $\my$ or high $\xpom$ ($\my>5\,{\rm GeV}$
or $\xpom>0.1$) is modelled with DJANGO~\cite{django}. This model is a
deep-inelastic $ep$ generator based on the QCD substructure of the
proton, with a colour string between the proton remnant and the
current fragmentation region.  Higher order QCD corrections are
provided by the colour dipole model and QED radiative effects are
implemented via an interface to HERACLES.  DJANGO has been shown to
reproduce well many aspects of the DIS final state including
measurements of energy flows in the forward region~\cite{H1eflow}.
The normalisation for DJANGO was chosen such that the sum of the cross
sections from RAPGAP, DIFFVM and DJANGO is in approximate agreement
with that corresponding to the measured proton structure function
$F_2(x,Q^2)$~\cite{f2paper}.


The PHOJET~\cite{phojet} generator was used to estimate background from
photoproduction events. In PHOJET events are generated as five distinct
subprocesses: non-diffraction, elastic vector meson production, vector meson
production with proton dissociation, single photon diffraction dissociation and
double diffraction dissociation.  These subprocesses are mixed such that the
total photoproduction cross section and the subprocess ratios are in
agreement with measurements made at HERA~\cite{totgammap}.

Background events arising from the QED Compton process ($ep \rightarrow
ep \gamma$) were simulated using the COMPOS generator~\cite{compos}.




\section{Experimental Technique}
\label{sec:tech}
The data presented here were obtained by H1 in 1994, when HERA was colliding
$27.5\,{\rm GeV}$ positrons with protons at $820\,{\rm GeV}$. The results
shown are based on an integrated luminosity of $1.96\,{\rm pb^{-1}}$ with the
interaction point in its usual position relative to H1. A further $0.06\,{\rm
  pb^{-1}}$ of data, taken with the interaction point shifted by $70\,{\rm
  cm}$ in the proton direction, are also used. These data enable measurements
to be made at lower values of $Q^2$. The two event samples are referred to as
the nominal vertex and shifted vertex data respectively.
 
\subsection{H1 Apparatus}
\label{sec:H1exp}
The H1 detector is a composite, multi-purpose apparatus, designed to
investigate the final states of high energy electron-proton
interactions. Given here is only a brief outline of the components of
the H1 detector of relevance to the present analysis; for a more
detailed technical description see~\cite{H1:detector}. A right handed
co-ordinate system is employed at H1 that has its $z$-axis pointing
in the proton beam, or forward, direction.

Surrounding the interaction point are the central and forward tracking
detectors. These are situated in a uniform $1.15\,{\rm T}$ magnetic
field, enabling momentum measurement of charged particles over the
pseudorapidity range $-1.5< \eta<2.8$.\footnote{The pseudorapidity
  ($\eta=- \ln \tan \theta/2$) coverage of each detector component is
  given for the vertex in its nominal position.} A multi-wire
proportional chamber (BPC) gives additional acceptance for charged
particles in the backward region ($-3.0<\eta<-1.5$).  Surrounding the
track detectors in the forward and central directions ($-1.5<\eta<3.7$) is a
fine grained liquid argon calorimeter (LAr) and in the backward region
($-3.5<\eta<-1.5$) is a lead-scintillator electromagnetic calorimeter
(BEMC).  The main calorimeters are surrounded by an instrumented iron
backing calorimeter.

Components of the H1 detector positioned forward of the LAr are used in
this analysis to tag particles at high values of pseudorapidity. These
`forward detectors' are sensitive to the secondaries produced by the
rescattering of primary particles with the beam pipe and adjacent material,
and are thus sensitive to larger pseudorapidities than would be expected on
the basis of their geometrical acceptance alone. A copper-silicon plug
calorimeter (PLUG) allows energy measurements to be made over the range
$3.5<\eta<5.5$. Three double layers of drift chambers of the forward muon
detector (FMD) are sensitive to particles produced at higher pseudorapidities
$5.0<\eta<6.5$. The proton remnant tagger (PRT), consisting of seven double
layers of scintillators, covers the region $6.0<\eta<7.5$.

Two electromagnetic calorimeters (LUMI) situated downstream in the electron
beam direction measure electrons and photons from the bremsstrahlung 
process $ep\rightarrow
ep\gamma$ for the purpose of luminosity determination. 


\subsection{Event Selection}
\label{sec:eventsel}
Deep-inelastic scattering events are selected by requiring a compact
energy deposit of at least $10.5\,{\rm GeV}$ in the BEMC and a BPC hit
associated with this cluster. Background in the DIS sample is
suppressed by requiring a vertex reconstructed by the track detectors
within $30\,{\rm cm}$ of the  mean $z$ position of the vertex, a
minimum $Q^2$ and at least two reconstructed non-associated objects
(tracks or clusters) besides the scattered electron in the final
state.  Further cuts, requiring an agreement between different
kinematic reconstruction methods, are introduced to reduce radiative
corrections.  The precise cuts employed are summarised in
table~\ref{tab:electronid}a. A more extensive discussion of the DIS
event selection at H1 may be found in~\cite{f2paper}.

\begin{table}[h]
\begin{center}
a) \,
\begin{tabular}{|c|c|} \hline
(1) & $ E_e^\prime > 10.5\,{\rm GeV}$    \\ \hline
(2) & $ Q^2 > 8.0$ $(3.5)\,{\rm GeV^2}$      \\ \hline
(3) & $156^\circ (163^\circ)< \theta_e <174^\circ (176^\circ)$   \\ \hline 
(4) & $\varepsilon_1 <5\,{\rm cm}$  \\ \hline
(5) & $\varepsilon_2 <5\,{\rm cm}$ \\ \hline
(6) & $|z_{\rm vtx}-\bar{z}_{\rm vtx}|<30\,{\rm cm}$ \\ \hline
(7) & $N_{\rm track}+N_{\rm cluster}>1$ \\ \hline
(8) & $-0.55<\yjb-y_{_e} <0.25$ \\ \hline
(9) & $|\yda-y_{_e}|<0.2 (0.3)$ \\ \hline
\end{tabular}
\,\,\,\,\,\,\,\,\,
b)
\,
\begin{tabular}{|c|c|} \hline
(10) & $\etamax<3.3$ \\ \hline
(11) & $E_{\rm PLUG}<5\,{\rm GeV}$ \\ \hline
(12) & $N_{\rm FMD}<2$ \\ \hline
(13) & $N_{\rm PRT}=0$ \\ \hline
\end{tabular}
\end{center}
\scaption{Summary of event selection criteria for a) inclusive DIS events and
  b) additional cuts to select large rapidity gap events. The shifted vertex
  cuts are given in brackets where they are different from the nominal vertex
  cuts. Section~\ref{sec:kinrec} gives details of the methods used to
  reconstruct the DIS kinematic variables. Cuts 1--3 ensure the scattered
  electron is within the acceptance region of the BPC and BEMC; cuts 4 and 5
  are electron identification estimators, $\varepsilon_1$ is the electron
  cluster radius, $\varepsilon_2$ is the distance from the closest hit in the
  BPC to the centroid of the electron cluster~\cite{f2paper}; cuts 6 and 7
  require there to be at least 2 reconstructed objects in the hadronic final
  state of which one must be a track pointing to the interaction region; cuts
  8 and 9 ensure agreement between different kinematic reconstruction
  techniques; cut 10 requires that there is no activity observed in LAr above
  noise levels for $\eta>3.3$; cuts 11--13 ensure that there is no activity
  above noise levels in the forward detectors~\cite{andythesis}.}
\label{tab:electronid}
\end{table}

A subsample of events is selected where colourless exchange processes (such
as diffraction) are expected to dominate.  This is achieved by requiring no
activity above noise levels in any of the forward detectors or in the most
forward part of the LAr (see table~\ref{tab:electronid}b for details of the
cuts).  The selection implies that systems $X$ and $Y$ (as defined in
section~\ref{sec:form}) are well separated, such that $X$ is fully contained
in the main detector components and $Y$ passes unobserved into the forward
beam pipe. The maximum allowed value of pseudorapidity for particles from the
system $X$ of 3.3 (cut (10) in table~\ref{tab:electronid}b) ensures that
$\xpom \lapprox 0.05$. The fact that no part of the system $Y$ is detected in
the forward detectors restricts $\my \lapprox 1.6\,{\rm GeV}$ and $\modt
\lapprox 1\,{\rm GeV^2}$.
This sample of
events is referred to hereafter as the `large rapidity gap' event
sample. The number of selected events is 13956 in the nominal vertex sample
and 760 in the shifted vertex sample.
 
The only backgrounds in the large rapidity gap event sample that were
found to be significant were from photoproduction and the QED Compton process
$ep \rightarrow ep\gamma$. Photoproduction background arises when the
scattered electron escapes undetected along the beam pipe and a final state
hadron fakes a BEMC electron cluster. This background was estimated using the
PHOJET simulation and is found to contribute most significantly at high $y$ and
low $Q^2$. The overall photoproduction background was estimated to be
0.3\,\% of the large rapidity gap event sample and at most 13\,\% for any
single data bin. 

QED Compton events were studied with the COMPOS generator and found to be
effectively suppressed by cut (7) in table~\ref{tab:electronid}. Residual QED
Compton background arises when the final state photon converts in material in
front of the BEMC, producing a BPC hit and is mistaken for the scattered
electron. The true electron and some residual calorimeter noise fake the
hadronic final state.  This background was found to contribute most
significantly at high $\beta$ and high $y$. The overall QED Compton
background is estimated to be 0.6\,\% of the large rapidity gap event sample
and contributes a maximum of 15\,\% in any single bin.

Each background source was subtracted bin by bin.


\subsection{Reconstruction of the Kinematic Variables}
\label{sec:kinrec}
The kinematic variables, defined in section~\ref{sec:form}, are calculated
from the energy and the polar angle of the scattered electron ($E_e^\prime$,
$\theta_e$) and from the 4-vector of the hadronic final state ($E_h$,
$p_{xh}$, $p_{yh}$, $p_{zh}$).  $E_e^\prime$ is reconstructed from the
cluster with the largest energy in the BEMC passing the cuts described in
table~\ref{tab:electronid}, and $\theta_e$ is the angle defined by the
$z$-vertex and the BPC hit associated with the electron.  The hadronic
4-vector is reconstructed by combining information from the track detectors
and calorimeters using an energy flow algorithm, which improves the
resolution compared to methods relying on single detector
components~\cite{mxpaper}.

\begin{figure}[htb]
\vspace{-1cm}
   \begin{center}
\epsfig{figure=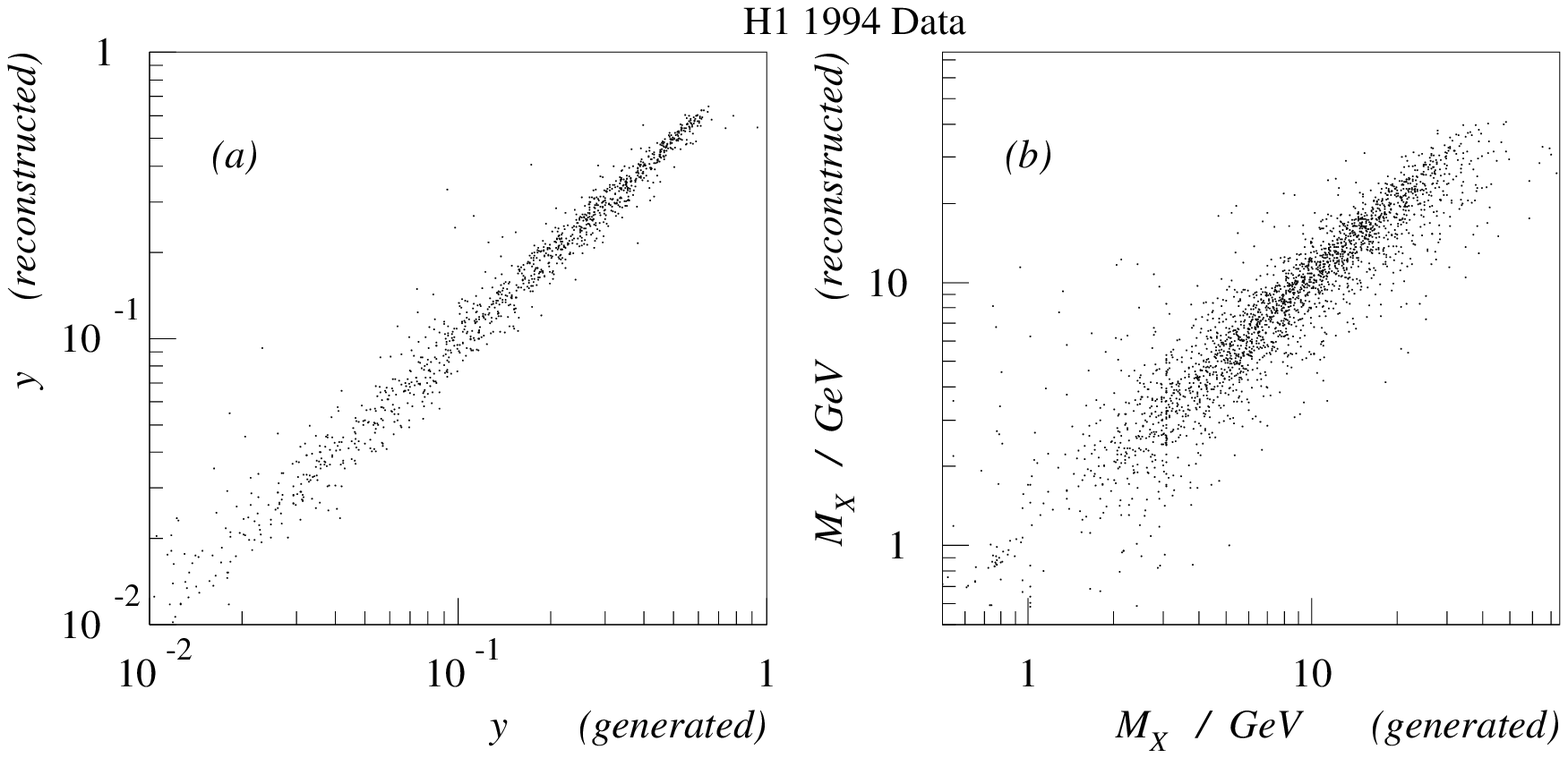,width=\textwidth}
   \end{center}
   \scaption{The correlation between the generated and reconstructed values
     of (a) $y$ (reconstructed using the weighted average method)
     and (b) $\mx$ for simulated Monte Carlo mixture at nominal
     vertex. }
\label{ymxgenrec}
\end{figure}

A key feature of the H1 experiment is its ability to measure both the
electron and hadronic final state well, thus providing a number of methods
for reconstructing the kinematic variables. In this analysis the kinematics
are reconstructed using a method that combines the electron  method, in
which the kinematics are reconstructed solely from the scattered electron,
and the double angle method, which uses only angles of the hadronic final
state and the scattered electron~\cite{doubleangle}. The two methods are
combined in order to utilise the optimal resolution of the electron method at
high $y$ and the double angle method at low $y$. We reconstruct $y$ as
\begin{equation}
y=y^2_e+\yda(1-\yda), 
\label{eqn:yweight}
\end{equation}
\begin{equation}
 {\rm where} \, \, \, \, \, \, \,
y_{_e}=1-\frac{E_e^\prime}{E_e} \sin^2(\theta_e/2) \,\,\, \,\,\, {\rm and} 
\,\,\, \,\,\, \yda= \frac{\sin \gamma (1+\cos \theta_e)}{\sin \gamma
  +\sin \theta_e+\sin (\theta_e+\gamma)},
\end{equation}
\begin{equation}
  {\rm with} \, \, \, \, \, \, \, \cos \gamma=\frac{p_{xh}^2+p_{yh}^2-(E_h-p_{zh})^2}{p_{xh}^2+p_{yh}^2+(E_h-p_{zh})^2},
\end{equation}
and $E_e$ is the electron beam energy. As demonstrated by
figure~\ref{ymxgenrec}(a), this `$y$ weighted averaging method' provides
excellent resolution in $y$ over its full range. $Q^2$ is reconstructed from
$y$ and the angle of the scattered electron
\begin{equation}
Q^2=\frac{4 E^2_e (1-y)}{\tan^2 (\theta_e/2)}.
\end{equation}

Two additional kinematic reconstruction methods are used to reconstruct $y$
and/or $Q^2$ for the purposes of cross checks and event selection cuts. These
are the hadron  method~\cite{jacquetblondel} for which
$\yjb=(E_h-p_{zh})/(2E_e)$ and the $\Sigma$ method~\cite{sigma} for which
$\ysig=\yjb/(1+\yjb-y_{_e})$ and $\qssig=4 E_e^{\prime 2}
\sin^2(\theta_e)/(1-\ysig)$.  The Bjorken scaling variable $x$ is
calculated in all methods as $Q^2/sy$ where $s$ is the $ep$ centre of mass
energy.

For the large rapidity gap event sample the system $X$ is fully contained
within the main detector, so that the visible hadronic final state
corresponds to the system $X$. It is thus possible to reconstruct the
variables $\mx$, $\beta$ and $\xpom$ well. The invariant mass $\mx$ is
reconstructed with a method that utilises the optimised reconstruction of $y$
\begin{equation}
  \mx^2= {(E_h^2-p^2_{xh}-p^2_{yh}-p^2_{zh})} \frac{y}{\yjb}.
\label{mxrec}
\end{equation}
A constant scale factor of 1.10, determined from the simulations, is applied
to the reconstructed $\mx$ to correct for residual losses, mainly due to dead
material in front of the calorimeters.  As demonstrated in
figure~\ref{ymxgenrec}(b), the resulting $\mx$ is well reconstructed across
the kinematic range of the measurement with a resolution of about 25\,\%.
The other kinematic quantities are determined using
equations~(\ref{eq:betaQ2}) and~(\ref{eq:xpomQ2}), with the approximation
$\modt\ll \mx^2+Q^2$. 

\begin{figure}[b]
   \begin{center}
\epsfig{figure=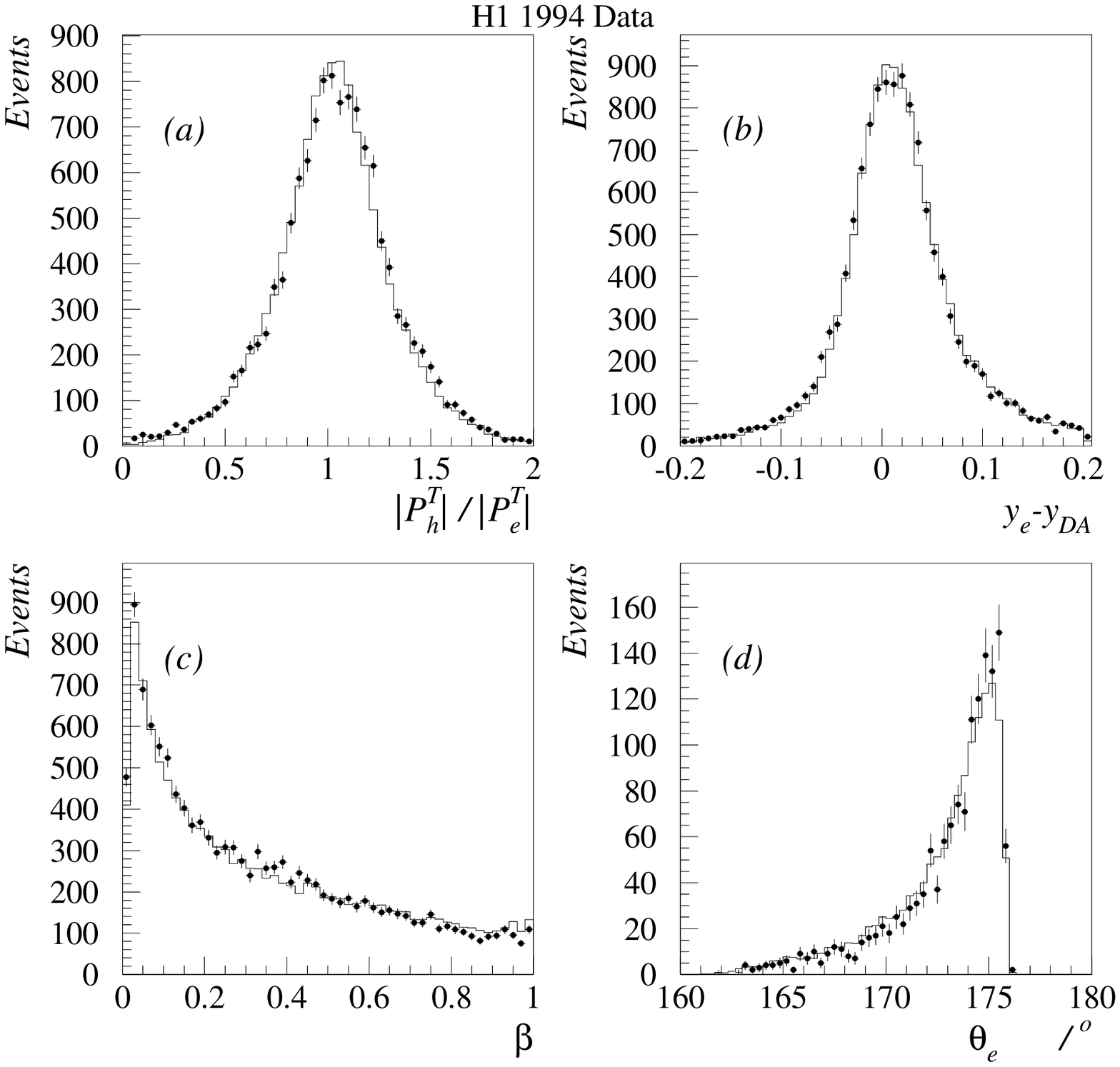,width=1.1\textwidth}
   \end{center}
   \scaption{The solid points show (a) the ratio of the transverse momentum
     of the hadrons to that of the electron, (b) the difference between $y$
     calculated from the electron and double angle methods, (c) the $\beta$
     distribution and (d) the electron polar angle distribution. All
     distributions are shown for the rapidity gap DIS event
     sample. Plots (a), (b) and (c) are for the nominal vertex data and plot
     (d) is for the shifted vertex data. None of the distributions have been
     corrected for detector effects. The predictions from the simulation are
     shown for each plot as a solid histogram.}
\label{control}
\end{figure}

The accuracy of the reconstruction method and the ability of the simulation
to describe the data were checked in a variety of ways.
Figure~\ref{control}(a) shows the ratio of the transverse momentum of the
visible hadrons to that of the electron, which for a perfect detector should
have a narrow width and peak at 1.  At the detector level it can be seen that
the peak is smeared with $\sigma \approx 30\%$, but with a mean close to 1,
showing that there are no substantial losses of final state hadrons.
Figure~\ref{control}(b) shows that there is good agreement between $y$ when
reconstructed using the electron or double angle methods before combination.
For both these distributions the simulation is in good agreement with the
data, showing that the resolutions and biases of the detector are well
understood.

Since an understanding of bin migrations depends not only on the resolution
in the kinematic quantities, but also on the form of the differential cross
section, it is important that the cross section used as the input to the
simulations closely matches that of the data.  Figures~\ref{control}(c) and
\ref{control}(d), which show the reconstructed $\beta$ and $\theta_e$
spectra, give an indication that this is indeed the case. Similar agreement
is found for the $\xpom$, $\mx$ and $y$ spectra.

\subsection{Evaluation of the Differential Structure Function}
\label{f2d}
The large rapidity gap event sample is used to measure the
differential cross section $\dthreesigb$ for the process $e p
\rightarrow e X Y$ as defined in section~\ref{sec:form} for
$4.5<Q^2<75\,{\rm GeV}^2$, $0.0002<\xpom<0.04$ and $0.04<\beta<0.9$.
The data are divided into three-dimensional bins in $x$, $\beta$ and
$Q^2$. The Monte Carlo models are used to correct bin by bin for
losses, smearing and backgrounds. No difference was observed within
the statistical errors of the simulations between the acceptances when
evaluated using the ARIADNE or parton showers fragmentation options
within RAPGAP. The average of the two acceptances was therefore taken.
An additional overall correction of $1.11\pm 0.05$ is applied to
account for events lost due to noise in the forward detectors, which
is not simulated in the Monte Carlo models.  This correction is
evaluated from randomly triggered events, in which there is no
requirement of particle activity in the detector. Radiative
corrections are
applied bin by bin using the HERACLES interface to RAPGAP.

Since the cross section is expected to vary smoothly with the kinematic
variables for bins that lie away from the resonance region of the light
vector mesons ($\rho^0$, $\omega$, $\phi$, $\rho^\prime$), the cross section
is corrected to that at the bin centre ($x_c$, $\beta_c$, $Q^2_c$) for bins
where $M_{Xc}>2\,{\rm GeV}$.  For these bins the main deviation from a smooth
variation of the cross section is expected to arise from elastic $J/\psi$ and
$\psi(2S)$ production, which contribute a maximum to any single bin of
approximately 5\% and 1\% respectively~\cite{pierre,vicki}. These
contributions are not subtracted but a systematic error equal to the expected
magnitude of the contribution is added to the systematic error.  For the bins
that have a mass interval such that they cover one or more of the light
vector mesons ($M_{Xc}<2\,{\rm GeV}$), the dependence of the cross section on
$x$ and $Q^2$ is expected to be smooth, but there are likely to be
substantial deviations from a smooth dependence of the cross section as a
function of $\beta$. For this reason the cross section is corrected to the
bin centre only in ($x_c$, $Q^2_c$) and in $\beta$ it corresponds to the
average over the range covered by the bin.

For all selected bins the acceptance is required to be above 0.4, and the
purity (i.e. the fraction of the number of events both generated and
reconstructed in a bin to the number reconstructed in the bin) is required to
be greater than 0.2. The chosen bins have an average acceptance of $0.82$ and
an average purity of $0.45$.  The two data samples are combined by taking the
shifted vertex data where there is no nominal vertex coverage ($Q^2<9\,{\rm
  GeV}^2$)
and the nominal vertex data elsewhere ($Q^2 \ge 9\,{\rm GeV}^2$). 

The statistical errors are formed by adding in quadrature the
contribution from the data and that due to the finite statistics of
the Monte Carlo simulations.  Systematic errors arise from the
following sources:

\begin{itemize}
\item The error due to uncertainties in the reconstruction of the scattered
  electron, estimated by changing the electromagnetic energy scale of the
  BEMC by $\pm 1\%$ and shifting the electron polar angle by $\pm 1\,{\rm
    mrad}$.
\item Uncertainties in the reconstruction of the hadronic final state,
  estimated by changing the hadronic energy scale of LAr by $\pm 5\%$, that
  of the BEMC by $\pm 20\%$ and by shifting the energy fraction carried by
  the tracks by $\pm 3\%$.
\item Errors due to our {\it a priori} ignorance of the input structure
  function as used in the simulations, estimated by reweighting the generated
  events by factors $\xpom^{\pm 0.2}$ and $(1 \pm 0.3\beta)$. These factors
  are chosen such that resulting generated structure function is changed by
  an amount larger than the final errors on the data points. An additional
  error is assigned to account for smearing of events from kinematic regions
  where the differential structure function is not measured. This error is
  conservatively estimated by changing the number of events generated with
  $\xpom>0.1$ and/or $\my>5\,{\rm GeV}$ by $\pm 50\%$.
\item Errors due to the uncertainty in the $t$ dependence, estimated by
  reweighting by ${e^{\pm 2 t}}$.
\item The uncertainty in the correction due to smearing of events across the
  boundary $\my=1.6 \,{\rm GeV}$, estimated by changing the PLUG energy scale
  by $\pm 30\%$, by changing the PRT efficiency by $\pm 20\%$, by taking the
  difference between the correction evaluated using isotropic and limited
  $p_T$ proton dissociation models and by varying the ratio of double
  dissociation to single photon dissociation assumed in the simulation
  between 0.5 and 2.0, such that the range covers that measured in various
  proton dissociation processes in DIS and
  photoproduction~\cite{mxpaper,h1phi,h1jpsi}.
\item The uncertainty in the number of background events,
 estimated as $50\%$ of the photoproduction and $30\%$ of the
  QED Compton  background.
\item The uncertainty in the bin centre correction due to contributions from
  high mass resonances, estimated by assuming an error of 5\% for all bins
  which contain the $J/\psi$ and 1\% for all other bins to account for the
  $\psi(2S)$ and other possible contributions.
\item An error of $2\%$ per point is assumed to account for any detector
  inefficiencies not included in the simulations~\cite{f2paper}.
\item An overall normalisation uncertainty of $4.8\%$ for the nominal vertex
  data and $6.0\%$ for shifted vertex data, evaluated by adding in quadrature
  a $4.5\%$ error on the uncertainty due to the correction for events lost
  due to forward detector noise and the errors on the integrated luminosity
  of $1.5\%$ and $3.9\%$ for nominal and shifted vertex data.
\end{itemize}

\begin{figure}[t]
   \begin{center}
     \vspace{-2cm} \epsfig{figure=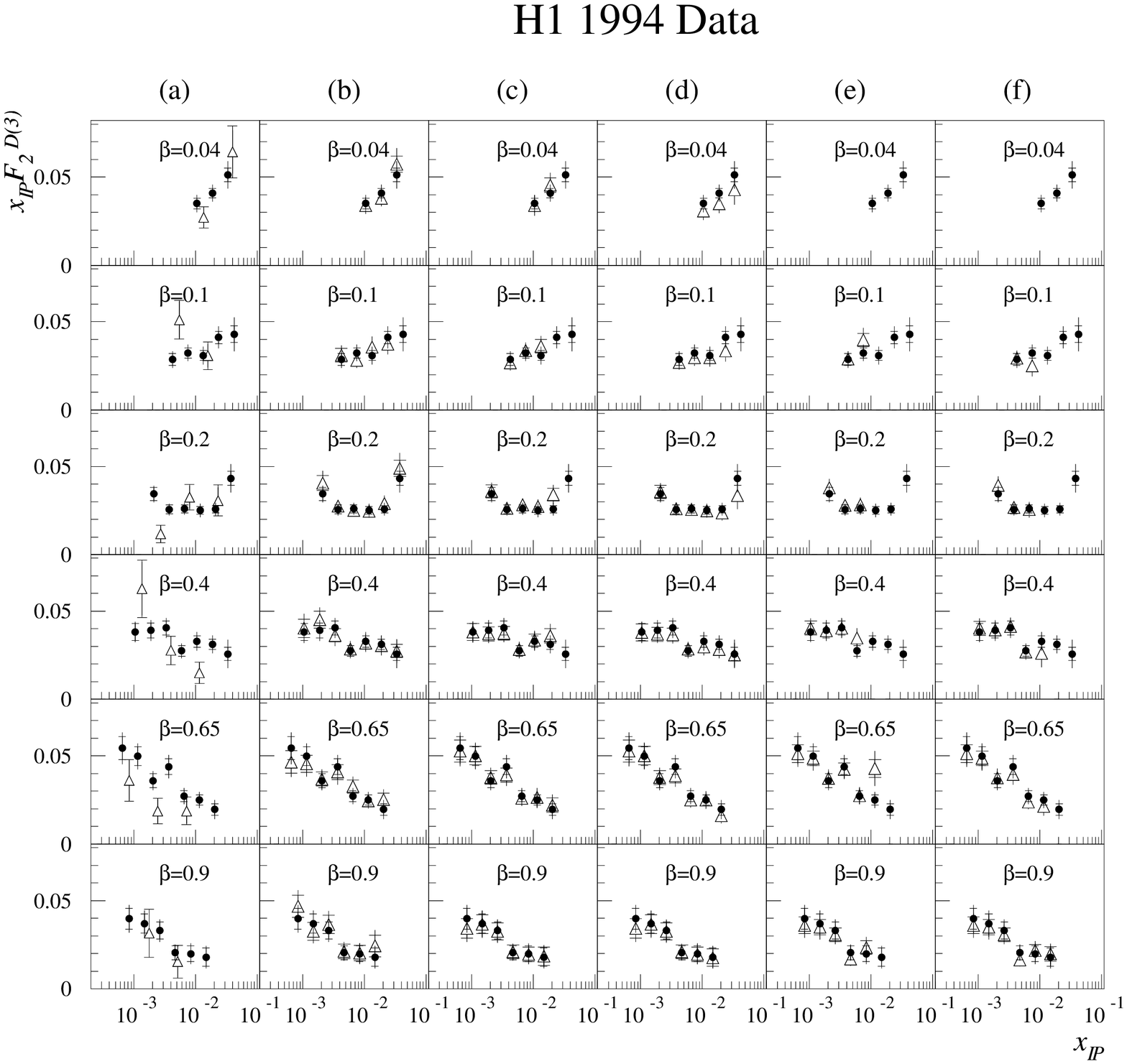,width=1.1\textwidth}
   \end{center}

   \scaption{Effect of changing kinematic reconstruction techniques and
     rapidity gap selection criteria on the measured values
     of $\xpom \ftdtb$ at an example
     $Q^2$ value of $18\,{\rm GeV^2}$.  The solid circles show the data
     measured using the standard analysis. In each of the six columns, these
     data are compared to those obtained with a modified analysis technique
     in which, (a) the shifted vertex data are compared (b) $x$ and $Q^2$
     were reconstructed with the $\Sigma$ method and $\mx$ solely from the
     hadronic final state, (c) cut 10 (see table ~\ref{tab:electronid}) is
     changed to $\etamax<3$, (d) cut 10 is not used, (e) cut 10 is replaced
     by $\etamax<2$ and there is no requirement on the forward detectors
     (i.e.  cuts 11--13 are not applied), (f) there is no rapidity gap
     selection at all (i.e. cuts 10--13 are not applied). The comparisons are
     plotted only for those bins satisfying the minimum acceptance and
     purity limits as for the standard analysis.}
\label{f2dcomp}
\end{figure}

To check the stability of the results, the analysis was repeated by widely
varying the experimental cuts and by changing kinematic reconstruction
techniques.  Figure~\ref{f2dcomp}(a) shows that there is reasonable agreement
between the results obtained from the nominal and shifted vertex data samples
for the standard analysis in the region of overlap. In
figure~\ref{f2dcomp}(b) it can be seen that the results do not change
significantly if the $\Sigma$ method is used to reconstruct the kinematics.
Figures~\ref{f2dcomp}(c), (d) and (e) demonstrate that changing the rapidity
gap requirement does not greatly change the results.  In
figures~\ref{f2dcomp}(f) it can be seen that even without the requirement of
a rapidity gap, when the measurement is made solely on the basis of the
reconstructed $\mx$ values in the inclusive data sample, the results obtained
are well within the quoted systematic uncertainty.

The measured values of $\ftdt$, evaluated using equation~(\ref{eq:F2D3}), are
listed in table~\ref{tab:results} and are also available 
from~\cite{QCD_Parameterisations}.

\begin{figure}[t]
   \vspace{-0.5cm}
   \begin{center}
     \epsfig{figure=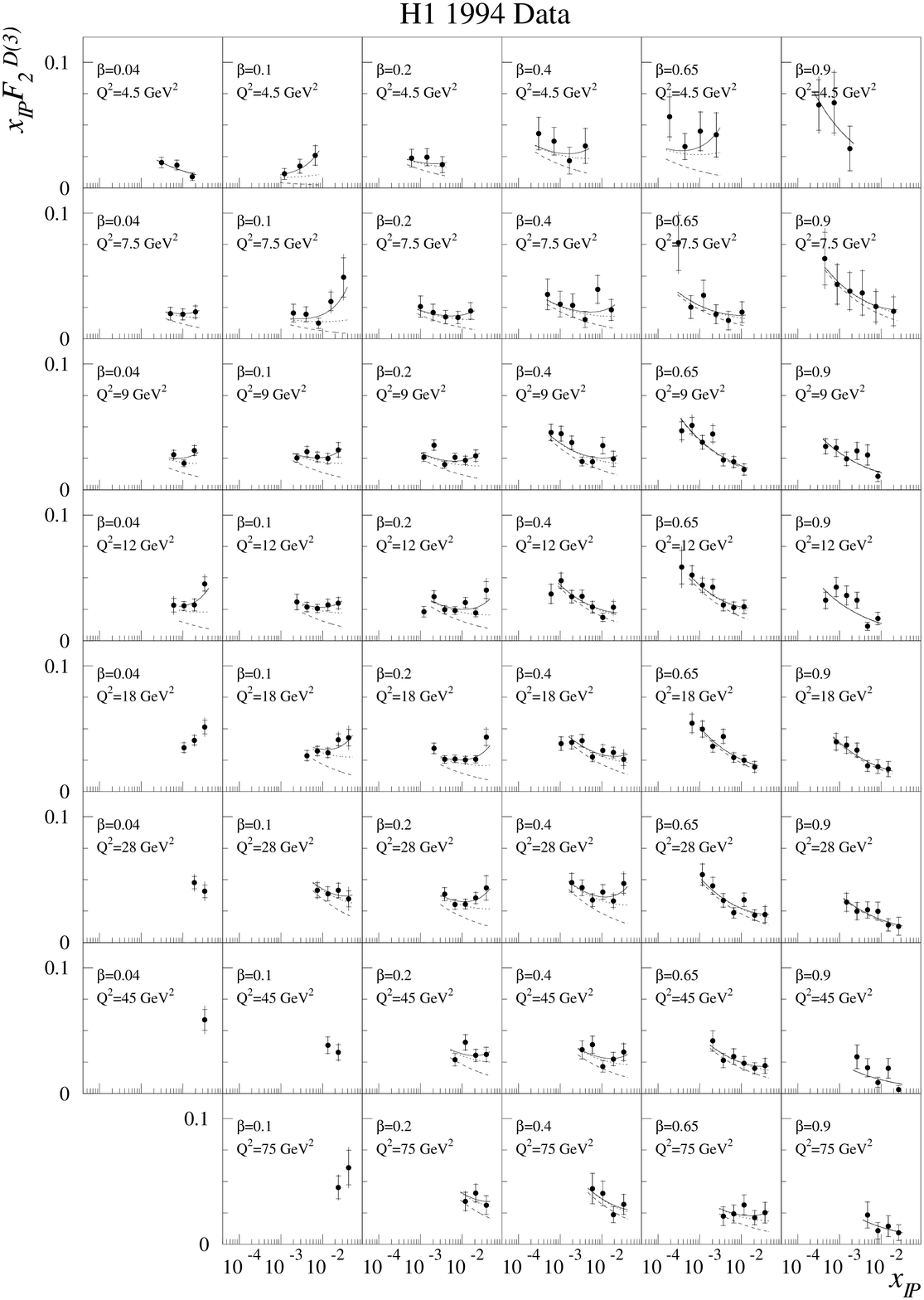,width=0.93\textwidth} \vspace{-0.3cm}
     \scaption{The solid points show the measured differential structure
       function plotted as $\xpom \ftdt$ against $\xpom$ for various $\beta$
       and $Q^2$ values. The inner error bars represent the statistical
       errors and the outer error bars represent the statistical and
       systematic errors added in quadrature.  The curves show the results of
       the Regge fit with interference (fit C) described in the text. The
       dashed curve is the contribution from the pomeron alone, the dotted
       curve is the pomeron plus interference and the continuous curve is the
       total contribution. The data that lie at values of $\xpom$ smaller
       than the extent of the curves have $y>0.45$ and are excluded from the
       fit.}
\label{fig:f2d}   
\end{center}
\vspace{-0.3cm}
\end{figure}

The quantity $\xpom \ftdt$ is shown\footnote{$\xpom \ftdt$ is plotted, rather
  than $\ftdt$, in order to show more clearly any deviations of the data from
  a simple power law behaviour on $\xpom$.} in figure~\ref{fig:f2d} as a
function of $\xpom$ for different $\beta$ and $Q^2$ values. It can be seen
that over most of the kinematic region covered $\xpom \ftdt$ is either
falling, or approximately constant, as function of increasing $\xpom$.
Qualitatively this is consistent with our earlier
observation~\cite{ref:h1_f2d_93} of a dominant diffractive contribution with
an intercept $\alphapom(0) \gapprox 1$.  However, the more precise data of
the present analysis show that there is a clear change in the $\xpom$
dependence at low $\beta$ where there is a tendency for it to increase at the
highest $\xpom$.  A quantitative study of this $\xpom$ behaviour is given in
terms of Regge amplitudes in the following section.


\section{Regge Parameterisation  of the Data}
\label{sec:phenfit}
An appropriate phenomenological framework to study the behaviour of
the DIS colour singlet exchange cross section is provided by Regge
theory~\cite{Regge,Chew}.  This theory has proved to be remarkably
successful in describing high energy hadron-hadron and hadron-photon
cross sections (see for
example~\cite{mxpaper,kaidalov:review,collins,DoLa}) as well as early
measurements of
$\ftdtb$~\cite{ref:h1_f2d_93,ref:zeus_f2d_93,zeus_LPS_94}.  Regge
theory in its simple form\footnote{No attempt has been made in this
  paper to describe $\ftdt$ using `triple-Regge' theory, 
  which would provide an additional prediction for the dependence of
  the cross section on $\beta$.}, when applied to deep-inelastic
scattering, provides information on the $\xpom$ dependence at fixed
$\beta$, $Q^2$ and $t$.  Since there is much theoretical uncertainty
concerning the precise form of the Regge model which is appropriate
for colour singlet exchange in DIS, several fits to the data are
performed with different assumptions. In all fits the
$\chi^2$ values are quoted for statistical errors only.

For fit A it is assumed that $\ftdtb$ may be factored into a pomeron
flux and pomeron structure function~\cite{Ingprytz}
\begin{equation}
\ftdt= \fpomnt  \ftpom.
\label{eqn:factorisation}
\end{equation}
The pomeron flux is taken to follow a Regge behaviour with a peripheral $t$
dependence and a linear trajectory $\alphapom(t)=\alphapom(0)+
\alphapom^{\prime} t$ such that
\begin{equation}
  \fpomnt=\int_{t_{cut}}^{t_{min}} \frac{e^{\bpom t}}{\xpom^{2\alphapom(t)-1}}
  {\rm d}t,
\end{equation}
where $|t_{min}|$ is the minimum kinematically allowed value of $|t|$ and
$t_{cut}=-1\, {\rm GeV}^2$ is the limit of the measurement.  The values for
$\bpom$ and $\alphapom^{\prime}$ cannot be constrained in the fit and are
assumed to be the same as those obtained from analyses of hadron-hadron data.
These values give consistent results with a measurement of the $t$ dependence
of the colour singlet exchange cross section averaged over the other
kinematic quantities~\cite{zeus_LPS_94}.  Table~\ref{tab:extraparams} lists
the assumed values for all fixed parameters used in this and subsequent fits.
The value of $\alphapom(0)$ is left as a free parameter in the fit as is
$\ftpom$ at each ($\beta$,$Q^2$) where there are data. A complication arises
from the lack of knowledge of $R^{D(5)}(\beta,Q^2,\xpom,t,\my)$, the ratio of
the longitudinal to the transverse cross section. The effect of a non-zero
value for $R^{D(5)}$ produces a decrease in the $\ftdtb$ extracted at large
$y$ relative to that obtained with the assumption $R^{D(5)}=0$. The maximum
value of $y$ for any of the H1 data is at $y=0.55$ which leads to a maximum
shift in $\ftdtb$ of $34\%$ if $R^{D(5)}=\infty$. Since these shifts are
large compared with the errors on the data points only those data with
$y<0.45$ are fitted thus limiting the shift to a maximum of $18\%$. The
difference when $R^{D(5)}=\infty$ instead of $R^{D(5)}=0$ is included as part
of the `model dependence' error for all fits.

\begin{table}[h]
\begin{center}
\begin{footnotesize}

\begin{tabular}{|c||lll||c|} \hline
 Quantity & \multicolumn{3}{c||}{Value} & Source \\ \hline 
   $R^{D(5)}$         & 0  & $^{+\infty}_{-0}$  & \hspace{-0.3cm}  &  \\ \hline 

$\alpha_{_{\pom}}^{\prime}$ & $0.26$ & $\pm 0.26$ & \hspace{-0.3cm} ${\rm GeV^{-2}}$ & \cite{cdf:elas
} \\ \hline 
$\alpha_{_{\reg}}^{\prime}$ & $0.90$ & $\pm 0.10$ & \hspace{-0.3cm} ${\rm GeV^{-2}}$ & \cite{apel:chexch} \\ \hline 
$\bpom$                & $4.6$  & $_{-2.6}^{+3.4}$  & \hspace{-0.3cm} ${\rm GeV^{-2}}$ & \cite{cdf:elas
} \\ \hline 
$\breg$                & $2.0$  & $\pm 2.0$  & \hspace{-0.3cm} ${\rm GeV^{-2}}$ & \cite{kaidalov
:review} \\ \hline 
\end{tabular}
\end{footnotesize}
\end{center}
\vspace{-0.3cm}
\scaption{Values assumed for the fixed parameters in the Regge fits and
the sources from which they are taken. Each quantity is defined in the text.
Contributions to the model dependence errors are formed by repeating the fits
after separately varying each parameter by the quoted uncertainties.}

\label{tab:extraparams}
\end{table}

Fit A yields a rather poor $\chi^2/{\rm ndf}$ of $257.8/168$.
Changing the form of the $\xpom$ dependence or the values of the fixed
parameters of the fit, whilst still retaining the factorisation
hypothesis, cannot explain the data simultaneously at both high and
low $\beta$.  The exchange of a single factorisable Regge trajectory
in the $t$-channel does not, therefore, provide an acceptable
description of the dependence of $\ftdtb$ on $\xpom$.

A natural explanation of the observed factorisation breaking is the
presence of a subleading exchange in addition to the leading pomeron.
Such an exchange is necessary to describe the energy dependence of
total, elastic and dissociative photoproduction and hadron-hadron
cross sections~\cite{kaidalov:review,mxpaper,dl:stot}.

In fits B and C we assume a subleading exchange, made up from the set
of reggeons which lie on the approximately degenerate trajectory
$\alphareg(t) \simeq 0.55+0.9 t$ and which carry the quantum numbers
of the $\rho$, $\omega$, $a$ or $f$ meson. It is assumed that each of
these exchanges can be expressed as the product of a flux and a meson
structure function akin to equation~(\ref{eqn:factorisation}) and that
the structure function of each exchange is the same. The situation is
complicated by the possibility of interference between the terms which
describe the pomeron and the $f$ meson
trajectory\footnote{Interference between the pomeron and any of the
  other subleading mesons trajectories is forbidden by C- and G-parity
  conservation.} so that $\ftdtb$ takes the form
\begin{eqnarray}
\ftdt= \fpomnt  \ftpom + \fregnt  \ftreg \nonumber \\
+ 2 I \fint \sqrt{\ftpom \ftreg},
\label{eqn:pomregfactorisation}
\end{eqnarray}
with
\begin{equation}
  \fregnt=\int_{t_{cut}}^{t_{min}} \frac{e^{\breg t}}{\xpom^{2 \alphareg(t)-1}}
  \ {\rm d}t.
\end{equation}
The interference flux is derived from the pomeron and meson fluxes 
with the phase completely specified by the signature factors of the pomeron
and the $f$~\cite{collins} in the form of
\begin{equation}
\fint=\int_{t_{cut}}^{t_{min}} \cos \left(\frac{\pi}{2}[\alphapom(t)-\alphareg(t)]\right) \frac{e^{(\bpom + \breg) t/2}}{\xpom^{\alphapom(t)+\alphareg(t)-1}}  {\rm d}t.
\end{equation}
The quantity $I$ denotes the degree of coherence and is assumed to take any
value from 0, when no $f$ contributes to the subleading exchange, and 
up to 1 for $f$ dominance of the subleading reggeon~\cite{mxpaper}.  In these
fits $I$ is assumed independent of $\beta$ and $Q^2$. The quantities
$\alphapom$, $\alphareg$, $\ftpom$ and $\ftreg$ are left as free fit
parameters and a constraint that $\ftreg\ge 0$ is imposed.


For fit B no interference is assumed by setting $I=0$.  The fit gives an
acceptable $\chi^2/{\rm ndf}$ of 120.7/121 and  describes the deviations
from factorisation of $\ftdtb$ well. Values of $\alphapom(0)=1.200\pm
0.017({\rm stat.})\pm 0.011({\rm sys.}) ^{+0.029}_{-0.034}({\rm model})$ and
$\alphareg(0)=0.57 \pm 0.14({\rm stat.})\pm 0.11({\rm sys.})
^{+0.03}_{-0.07}({\rm model})$ are obtained. The systematic errors on both
extracted quantities are estimated by adding in quadrature the differences
obtained from the central fit value after imposing each of the variations
described in section~\ref{f2d}.  The model dependence error is estimated in a
similar manner by varying the values of the fixed parameters of the fit over
the ranges listed in table~\ref{tab:extraparams}.  These ranges represent our
best estimate of the likely variation in the present data after examining the
spread in the values obtained in other scattering processes.

In fit C maximal interference is assumed between the pomeron and the meson
trajectory ($I=1$). A very similar $\chi^2/{\rm ndf}$ of 120.4/121 is
obtained, with a similar value of $\alphapom(0)=1.206\pm0.022({\rm stat.})\pm
0.013({\rm sys.}) ^{+0.030}_{-0.035} ({\rm model})$ and a lower value of
$\alphareg(0)=0.44\pm 0.08({\rm stat.})\pm 0.07({\rm sys.}) ^{+0.06}_{-0.05}
({\rm model})$.  The results of this fit are shown in figure~\ref{fig:f2d}.
It can be seen that at constant $\xpom$ the meson component and the
interference part fall much more steeply as $\beta$ increases than does the
pomeron component.

Since fits B and C both give a good description of the data and have similar
$\chi^2/{\rm ndf}$ it is not possible to determine, with the present data,
whether or not interference plays a significant role in DIS colour singlet
exchange cross sections. Further precision measurements at high values of
$\xpom$ are needed to clarify this matter. The results of the two fits do,
however, yield compatible results for the intercepts of the pomeron and meson
exchanges, and, since it is likely that the truth lies somewhere between the
solutions with $I=0$ and $I=1$, a best estimate for the values of the
intercepts is obtained by averaging the two results and including the 
difference between the two fits in the model dependent error. This procedure
yields
\begin{eqnarray*}
 \alphapom(0) & = & 1.203\pm 0.020 \ ({\rm stat.})\pm 
 0.013 \ ({\rm sys.})  ^{+0.030}_{-0.035} \ ({\rm model}) \\
\alphareg(0) & = & 0.50 \pm 0.11 \ ({\rm stat.})\pm 0.11
\ ({\rm sys.}) ^{+0.09}_{-0.10} \ ({\rm model}).
\end{eqnarray*}
The value of $\alphapom(0)$ found in this analysis is significantly larger
than that obtained from analyses of soft hadronic cross
sections~\cite{dl:stot,cudell}. It also exceeds the value
$\alphapom(0)=1.068\pm 0.016 ({\rm stat.}) \pm 0.022({\rm sys.})
\pm0.041({\rm model})$, which was obtained from an H1 analysis of the photon
diffractive dissociation cross section at $Q^2=0$~\cite{mxpaper}.  The value
of $\alphareg(0)$ agrees  with the value of $\simeq 0.55$ obtained in an
analysis of total hadronic cross sections~\cite{dl:stot}.

An attempt is made to see if there are any further deviations from
factorisation of the pomeron flux within the two-reggeon model described by
equation~(\ref{eqn:pomregfactorisation}).  Two separate fits are performed
that follow B, but now $\alphapom(0)$ is allowed to vary as a function of
either $\beta$ or $Q^2$ so that  $\alphapom(0)=a_1+a_2(\beta-0.5)$ or
$\alphapom(0)=a^\prime_1+a^\prime_2\log_{10}(Q^2/10{\rm GeV}^2)$. Both give
acceptable $\chi^2/{\rm ndf}$ of 120.5/120 and 120.7/120 respectively, but
within the errors both fits show that the data are compatible with no change
in $\alphapom(0)$ with $\beta$ or $Q^2$: $a_2=0.03 \pm 0.10({\rm stat.}) \pm
0.04 ({\rm sys.})^{+0.01}_{-0.06}({\rm model}) $ , $a^\prime_2=0.004 \pm
0.071 ({\rm stat.}) \pm 0.014 ({\rm sys.})^{+0.003}_{-0.003}({\rm model}) $.
Similar results are obtained for fits performed with full interference.

We conclude that the data may be described using a Regge model with
two components in which a pomeron contributes along with a subleading
exchange. In this two reggeon scenario there is, within experimental
errors, no evidence for a change in the pomeron intercept with $\beta$
or $Q^2$ over the kinematic region of this measurement.

\section{Deep-inelastic Structure of Diffractive Exchange}
\label{sec:qcdfit}
\subsection{QCD Extension of the Phenomenological Model}
In the universal factorisation hypothesis discussed in section
\ref{sec:phenfit}, the structure functions $F_2^{I\!\!P}(\beta,Q^2)$ and
$F_2^{I\!\!R}(\beta,Q^2)$ describe the deep-inelastic structure of the
pomeron and meson exchanges respectively. It has been suggested that
the $Q^2$ evolution of these structure functions may be understood in terms
of parton dynamics and therefore perturbative
QCD~\cite{IngSchlein,Ingprytz,TrentVenez}. 

To establish whether the data are consistent with such a hypothesis, the
phenomenological model presented in section \ref{sec:phenfit} is further
developed by assigning parton distribution functions to the pomeron and to
the meson. Currently, there is no clear theoretical consensus on how colour
singlet exchange processes may be understood within the context of QCD,
perturbative or otherwise~\cite{McDermott}. Therefore, a simple
prescription is adopted in which the parton distributions of both the pomeron
and the meson are parameterised in terms of non-perturbative input
distributions at some low scale $Q_0^2$. The parton distributions of the
pomeron and of the meson are evolved separately with increasing $Q^2$
according to the DGLAP~\cite{DGLAP} evolution equations.

For the pomeron, a quark flavour singlet distribution
($z{\cal{F}}_{q}(z,Q^2)=u\!+\!\bar{u}\!+\!d\!+\!\bar{d}\!+\!s\!+\!\bar{s}$)
and a gluon distribution ($z{\cal{F}}_g(z,Q^2)$) are parameterised in terms
of the coefficients $C_j^{i}$ at a
starting scale of $Q^2_0=3\gevt$ such that
\begin{equation}
z{\cal{F}}_{i}(z,Q^2=Q_0^2) = \left[
\sum_{j=1}^n C_j^{i} \cdot P_j(2z-1) \right]^2 
\cdot e^{\frac{a}{z-1}}
\label{eq:qcdparam}
\end{equation}
where $z=x_{i/I\!\!P}$ is the fractional momentum of the pomeron carried by
the struck parton. If the photon couples directly to a quark ($i=q$) then
$z=\beta$, whilst if the photon interacts with a gluon ($i=g$) then $z$ is the
fractional momentum carried by the gluon and $0<\beta<z$. $P_j(\zeta)$ is the
$j^{th}$ member in a set of Chebychev polynomials, which are chosen such that
$P_1=1$, $P_2=\zeta$ and $P_{j+1}(\zeta)=2\zeta P_{j}(\zeta)-P_{j-1}
(\zeta)$. A sum of $n$ orthonormal polynomials is used so that the input
distributions are free to adopt the widest possible range of forms for a
given number of parameters.  Any bias towards a particular solution due to
the choice of the functional form of the input distribution is therefore
minimised, and the number of terms included in the parameterisation may be
matched to the significance of the data.  The series is squared to ensure a
positive definite parameterisation for all values of $z$ and $C_j^{i}$.

For the evolution equations to be soluble, the parton distribution functions
must tend to zero as $z\rightarrow 1$. This is achieved by introducing the
exponential term with a positive value of the parameter $a$.  Unless
otherwise indicated, in the following fits $a$ is set to $0.01$ such that
this term only influences the parameterisation in the region $z>0.9$.

The functions $z{\cal{F}}_{i}(z,Q^2)$ are evolved to higher $Q^2$ using the
next-to-leading order DGLAP evolution equations and the contribution to
$F_2^{I\!\!P}(\beta,Q^2)$ from charm quarks is calculated in the fixed
flavour scheme using the photon-gluon fusion prescription given
in~\cite{GHRGRS}. The contribution from heavier quarks is neglected.

In the Regge model phenomenological fits  B and C described in
section~\ref{sec:phenfit} the fitted values of $\ftreg$ increase with
decreasing $\beta$ such that they are consistent in shape with
parameterisations of the pion structure function. Since the errors on
the current data are insensitive to any $Q^2$ evolution of $\ftreg$,
the structure function of the subleading exchange is taken from a
parameterisation of the pion~\cite{OWENS} multiplied by a single
coefficient $C_{I\!\!R}$ to be determined from the data.

No momentum sum rules are imposed because of the theoretical uncertainty in
specifying the normalisation of the pomeron or meson fluxes ($\fpomnt,
\fregnt$) and because it is not clear that such a sum rule is appropriate for
the parton distributions of a virtual exchange.

\subsection{QCD Fits of \boldmath{$\ftdt$}}
In the following fits the values of $F_2^{D(3)}(\beta,Q^2,\xpom)$ are
calculated from $F_2^{I\!\!P}(\beta,Q^2)$ and $F_2^{I\!\!R}(\beta,Q^2)$ using
equation~(\ref{eqn:pomregfactorisation}) assuming no interference between the
leading and subleading trajectories. All QCD fits are performed using the
total statistical errors. As in the phenomenological fits, the data for
$y>0.45$ are excluded from all of the QCD fits to limit the uncertainty
introduced by the lack of knowledge of $R^{D(5)}$. Data points corresponding
to values of $M_X$ less than $2\gev$ are also excluded because of the large
contribution to the cross section from the resonant production of vector
mesons and other possible higher twist contributions in this region.

\begin{figure}[t]
   \vspace{-2.5cm}
   \begin{center}
     \epsfig{figure=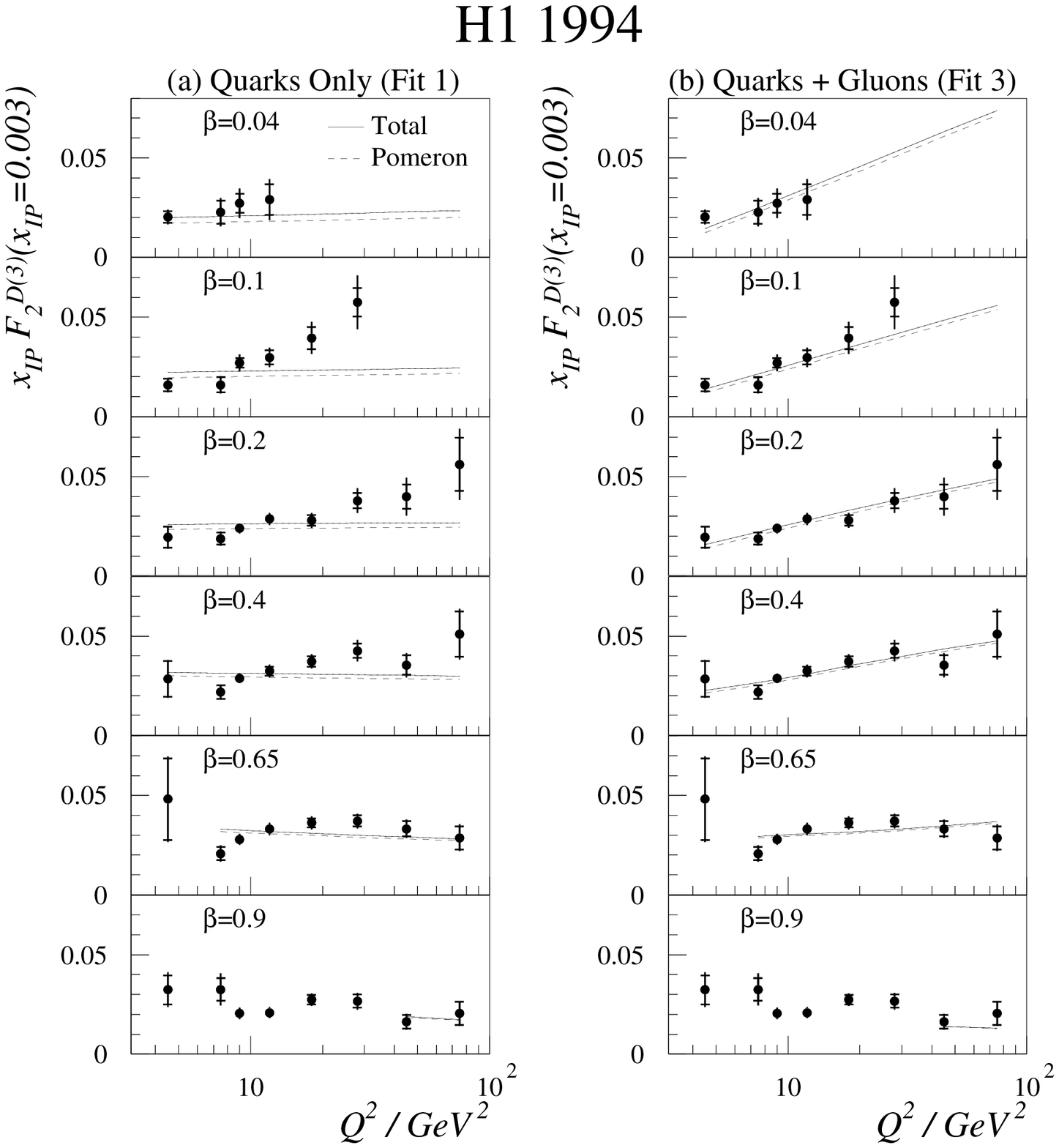,width=1.0\textwidth} \vspace{-0.3cm}
     \scaption{The quantity $\xpom\ftdt$ at $\xpom=0.003$ as a function
       of $Q^2$ for different values of $\beta$. In both (a) and (b) the solid
       points show the same data, interpolated to $\xpom=0.003$ using fit B
       (see text). In (a) the result of QCD fit 1 (in which only quarks
       contribute to the pomeron structure at $Q^2=3\gevt$) is superimposed.
       In (b) the preferred QCD fit 3 (in which both quarks and
       gluons contribute) is shown.  The results of the QCD fits are only
       shown in the kinematic range used in the fit (corresponding to data
       for which $M_X>2\gev$).  In both figures the
       solid line represents the value of $\ftdt$, whilst the dotted line
       shows the contribution from the pomeron term only.}
\label{fig:QCD1}   
\end{center}
\vspace{-0.3cm}
\end{figure}

\begin{figure}[t]
   \vspace{-2.5cm}
   \begin{center}
     \epsfig{figure=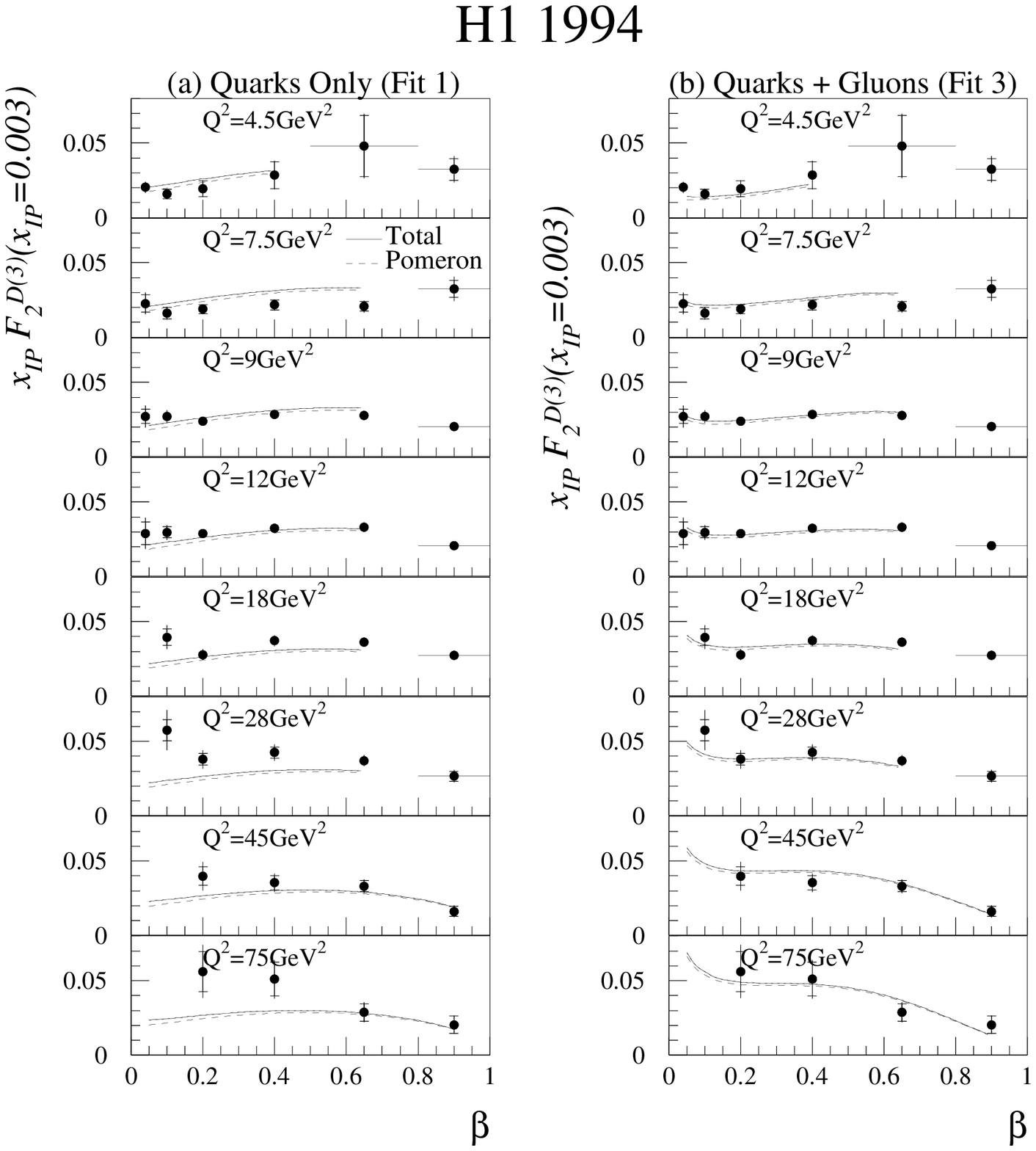,width=1.0\textwidth} \vspace{-0.3cm}
     \scaption{The quantity $\xpom\ftdt$ at $\xpom=0.003$ as a function
       of $\beta$ for different values of $Q^2$. In both (a) and (b) the solid
       points show the same data, interpolated to $\xpom=0.003$ using fit B
       (see text). In (a) the result of QCD fit 1 (in which only quarks
       contribute to the pomeron structure at $Q^2=3\gevt$) is superimposed. 
       In (b) the preferred QCD fit 3 (in which both quarks and
       gluons contribute) is shown.  The results of the QCD fits are only
       shown in the kinematic range used in the fit (corresponding to data
       for which $\mx>2\gev$).  In both figures the
       solid line represents the value of $\ftdt$, whilst the dotted line
       shows the contribution  from the pomeron term only.}
\label{fig:QCD1b}   
\end{center}
\vspace{-0.3cm}
\end{figure}

For fit $1$ only quarks are assumed to contribute to the structure of the
pomeron at the starting scale of $Q^2_0=3\gevt$, and the values of
$\alphapom(0)$ and $\alphareg(0)$ are fixed to those determined in the Regge
fit without interference (fit B).  The parameters to be fitted are then
$C_j^{q}$ and the normalisation of the subleading component $C_{I\!\!R}$.
Introducing successive terms from the Chebychev series, a $\chi^2/{\rm ndf}$
of $314.2/159$ is reached for $3$ terms. If an additional term is introduced
then the corresponding coefficient is consistent with $0$, with no
improvement of the quality of the fit.  The resulting parameterisation of
$F_2^{D(3)}$ is shown at a constant value of $\xpom=0.003$ as a function of
$Q^2$ for different values of $\beta$ in figure \ref{fig:QCD1}(a), and as a
function of $\beta$ at fixed $Q^2$ in figure \ref{fig:QCD1b}(a). Also shown are
the data for $\xpom F_2^{D(3)}$ interpolated to $\xpom=0.003$ using fit B.
The large $\chi^2/{\rm ndf}$ for fit 1 arises from the failure to reproduce
the scaling violations of $\ftdt$, which rises with increasing $Q^2$ for
$\beta<0.9$.  Therefore, the data cannot be described by a parameterisation
in which the pomeron contains only quarks at $Q^2=3\gevt$.

In fit 2 the first term in the expansion of the gluon distribution is
introduced as an additional fit parameter, and a significantly better
description of the data is obtained ($\chi^2/{\rm ndf}=186.7/156$). The
description of the data improves further with the introduction of the next
two terms in the series to give $\chi^2/{\rm ndf}=176.3/154$ for $n=3$ (fit
3). This fit is shown in figures \ref{fig:QCD1}(b) and \ref{fig:QCD1b}(b)
where it can be seen that the dependence of $\ftdt$ on $Q^2$ and $\beta$ is
well reproduced.  Introducing additional terms for the gluon distribution
does not result in any further improvement of the quality of the fit.

\begin{figure}[t]
   \vspace{-0.5cm}
   \begin{center}
     \epsfig{figure=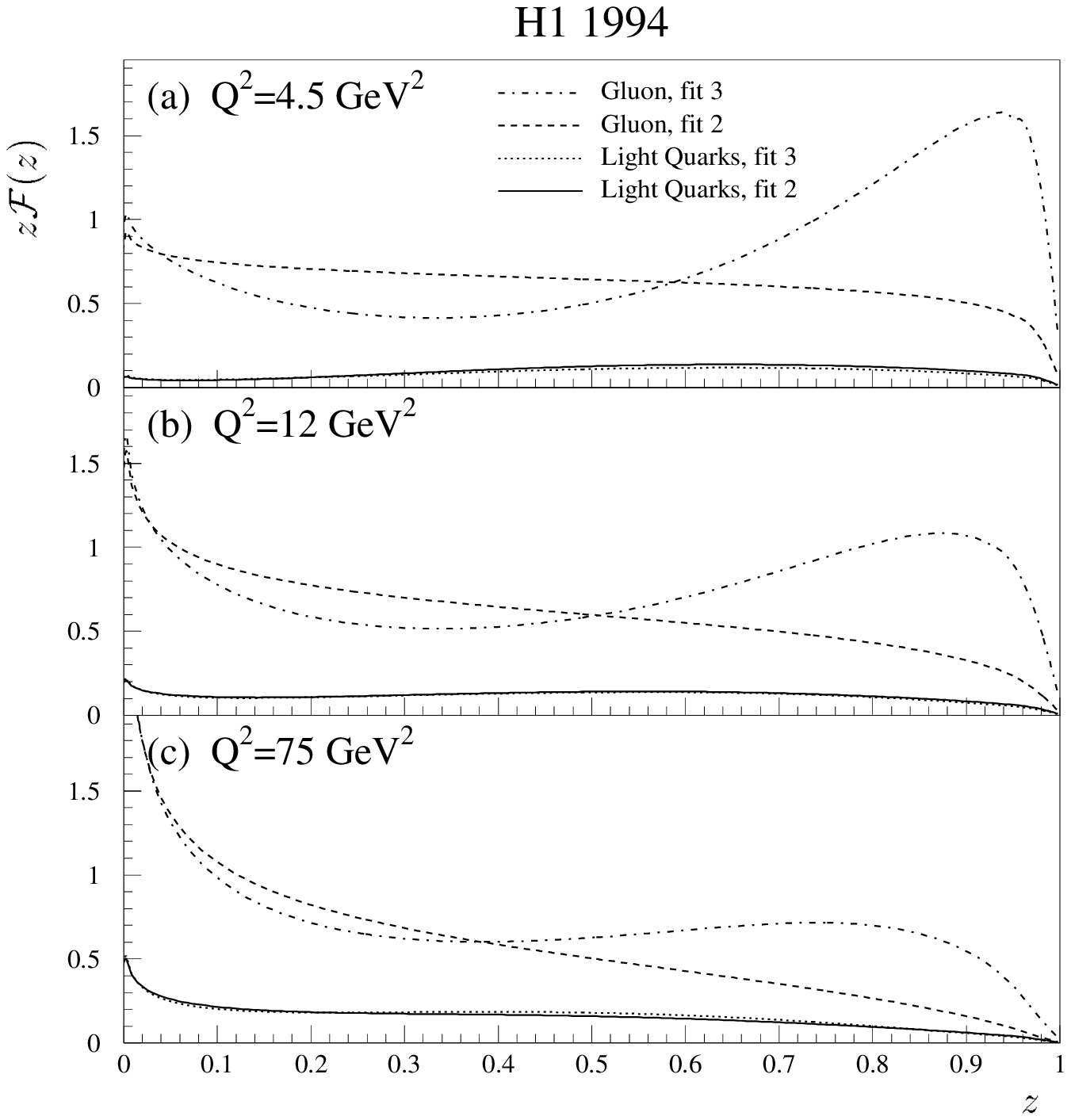,width=0.95\textwidth} \vspace{-0.3cm}
     \scaption{The sum of the light quark distributions and the gluon
       distribution for fits 2 and 3, shown at (a) $Q^2=4.5\gevt$, (b)
       $Q^2=12\gevt$ and (c) $Q^2=75\gevt$.  In fit 2 the gluon distribution
       is parameterised at the starting scale $Q_0^2=3\gevt$ with only the
       first term in the polynomial expansion, while in fit 3 the first 3
       terms are included. In both fits $3$ polynomials are used to
       parameterise the light quark distribution. The figures are
       normalised such that they represent the  parton distributions
       multiplied by the flux factor at $\xpom=0.003$.}
\label{fig:QCD2}   
\end{center}
\vspace{-0.3cm}
\end{figure}

The parton distributions resulting from fits $2$ and $3$ are shown in figure
\ref{fig:QCD2} where both the sum of the light quark distributions and the
gluon distribution are shown for each fit at the lowest and highest $Q^2$
values included in the analysis as well as an intermediate value of
$Q^2=12\gevt$.  In both fits 2 and 3 the fraction of the momentum of the
pomeron carried by gluons decreases with increasing $Q^2$ from $\sim 90\%$ at
$Q^2=4.5\gevt$ to $\sim 80\%$ at $Q^2=75\gevt$.  A large gluon component in
the pomeron is, therefore, favoured by the data.

Repeating fit 3 with $\alphapom(0)$ and $\alphareg(0)$ as free parameters
results in no change in either the coefficients $C_j^{q,g}$ or the quality of
the fit, and the values for $\alphapom(0)$ and $\alphareg(0)$ are unchanged
from those obtained in the Regge fit B. The parameters $a$ and $Q^2_0$ may
be varied in fit 3 within the ranges $0<a<0.1$ and $2<Q^2_0<4\gevt$ without
producing a change of more than one unit in the $\chi^2$, and if $a$ is
allowed to vary in the fit then the value obtained is consistent with $0.01$.
A final fit is performed to assess the influence of the particular choice of
the meson structure function by reweighting the parameterisation of
\cite{OWENS} by a factor $(1-\beta)^{\gamma_{I\!\!R}}$.  The value of
$\gamma_{I\!\!R}$ obtained is consistent with zero.

\subsection{Discussion of the QCD Fits}
We conclude that it is possible to describe the data for $F_2^{D(3)}$
in the region $M_X>2\gev$ with a factorisable model in which the
pomeron structure function evolves with $Q^2$ according to the NLO
DGLAP evolution equations.  Within the context of this model a large
gluonic content of the pomeron is required, with a hard distribution
in fractional momentum in order to achieve an adequate description of
the data.  The data exhibit some preference for a gluon distribution
that is large at high $z$ such that for low values of $Q^2$ the
majority of the momentum of the pomeron is carried by a single gluon,
although the precise shape of the gluon distribution is not well
constrained.  It should be noted, however, that there are some
unresolved theoretical problems associated with the use of the DGLAP
evolution equations at very large $z$ due to resummation
effects~\cite{resummation}, which may be particularly important when
the parton densities near $z=1$ are very large as in fit 3.

The fraction of the momentum of the pomeron carried by gluons is similar in
both fits 2 and 3, from $\sim 90\%$ at $Q^2=4.5\gevt$ to $\sim 80\%$ at
$Q^2=75\gevt$.  Thus the data lend support to pictures of diffraction in DIS
in which the dominant mechanism is boson-gluon fusion with the incoming gluon
carrying a large fraction of the momentum of the
pomeron~\cite{buchmueller,white}. 

Measurements of the properties of hadronic final states produced in
diffractive hard interactions are essential to clarify the extent to which
this partonic interpretation may be applied universally.  For example, a
large gluon distribution in the pomeron, peaked near $z=1$, was found to
describe well the measurements of diffractive dijet production in $p\bar{p}$
interactions~\cite{UA8}. The large gluon distribution favoured by the present
analysis predicts a sizable cross section for diffractive dijet and charm
production in both $pp$ and $ep$ interactions.  To facilitate comparisons
between such measurements~\cite{pphads} and the parton distribution functions
presented here, parameterisations of the parton distributions functions for
the NLO fits $1$, $2$ and $3$ are available~\cite{QCD_Parameterisations}.
Parameterisations are also available of the results of three leading order
QCD fits which are identical to the NLO fits $1$, $2$ and $3$ in all other
respects. The results of these additional fits, not shown here, lead to the
same conclusions as those outlined above.

The QCD analysis of $\ftdtb$ presented in this paper does not attempt to 
address the physical origin within QCD of either the intercepts of the 
leading and subleading exchanges, or of the parton distributions at the 
starting scale of the perturbative evolution. It is hoped that a more
complete understanding of colour singlet exchange interactions within QCD 
will establish the extent to which the DGLAP evolution prescription is 
appropriate, and also provide predictions for $\alphapom(0)$
and $\alphareg(0)$.

\section{Summary}
The differential structure function $\ftdt$ has been measured using the data
taken in 1994 with the H1 detector with greater accuracy and over a greater
kinematic range than previously.  At this level of precision it is found that
$\ftdtb$ is inconsistent with a universal dependence on $\xpom$. A natural
explanation of this breaking of factorisation is presented in which a leading
pomeron exchange is accompanied by a subleading meson exchange. Fits to the
data in which such a subleading term is considered yield an intercept for the
pomeron of $\apres$, which is significantly higher than the value obtained in
similar parameterisations of soft hadronic cross sections. No evidence is
found for any variation of $\alphapom(0)$ with $Q^2$ or $\beta$ in the range
$4.5<Q^2<75\,{\rm GeV^2}$ and $0.04<\beta<0.9$. The value obtained for the
intercept for the subleading exchange, $\amres$, was found to agree well with
values obtained from soft hadronic scattering processes.

A model in which the deep-inelastic structure of the leading exchange is
described by parton distributions which evolve according to the NLO DGLAP
evolution equations is consistent with the data for $\ftdt$ in the region
$M_X>2\gev$. Under such a hypothesis, the data require approximately $90\%$
and $80\%$ of the momentum of the pomeron to be carried by gluons at
$Q^2=4.5\gevt$ and $Q^2=75\gevt$ respectively. A parametrisation which
assumes only quarks in the diffractive exchange at $Q^2=3\,{\rm GeV^2}$ is
excluded.

 \section*{Acknowledgements}

 We are grateful to the HERA machine group whose outstanding
 efforts have made and continue to make this experiment possible. We thank
 the engineers and technicians for their work in constructing and now
 maintaining the H1 detector, our funding agencies for financial support, the
 DESY technical staff for continual assistance, and the DESY directorate 
 for the hospitality which they extend to the non-DESY members of the 
 collaboration. We thank J.~Bartels and T.~Gehrmann  for useful discussions.


\begin{table}[h]
\begin{scriptsize}
{\begin{tabular}{|c|c|c|c|c|c|} \hline
 $Q^2$ & $\beta$ & $x$ & $\xpom F_2^{D(3)}$ &$\delta_{\it stat}$ & $\delta_{\it sys}$  \\ \hline
 4.5 & 0.04 & 0.00012 & 0.0203 & 0.0042 & 0.0017 \\ \hline
 4.5 & 0.04 & 0.00029 & 0.0182 & 0.0039 & 0.0017 \\ \hline
 4.5 & 0.04 & 0.00067 & 0.0091 & 0.0030 & 0.0017 \\ \hline
 4.5 & 0.10 & 0.00012 & 0.0112 & 0.0044 & 0.0015 \\ \hline
 4.5 & 0.10 & 0.00029 & 0.0174 & 0.0055 & 0.0025 \\ \hline
 4.5 & 0.10 & 0.00067 & 0.0259 & 0.0077 & 0.0038 \\ \hline
 4.5 & 0.20 & 0.00012 & 0.0238 & 0.0070 & 0.0022 \\ \hline
 4.5 & 0.20 & 0.00029 & 0.0246 & 0.0068 & 0.0021 \\ \hline
 4.5 & 0.20 & 0.00067 & 0.0185 & 0.0063 & 0.0024 \\ \hline
 4.5 & 0.40 & 0.00012 & 0.0433 & 0.0130 & 0.0033 \\ \hline
 4.5 & 0.40 & 0.00029 & 0.0372 & 0.0110 & 0.0015 \\ \hline
 4.5 & 0.40 & 0.00067 & 0.0218 & 0.0107 & 0.0022 \\ \hline
 4.5 & 0.40 & 0.00160 & 0.0335 & 0.0139 & 0.0039 \\ \hline
 4.5 & $0.65^*$ & 0.00012 & 0.0566 & 0.0159 & 0.0112 \\ \hline
 4.5 & $0.65^*$ & 0.00029 & 0.0330 & 0.0104 & 0.0032 \\ \hline
 4.5 & $0.65^*$ & 0.00067 & 0.0451 & 0.0154 & 0.0055 \\ \hline
 4.5 & $0.65^*$ & 0.00160 & 0.0423 & 0.0175 & 0.0049 \\ \hline
 4.5 & $0.90^*$ & 0.00029 & 0.0660 & 0.0200 & 0.0105 \\ \hline
 4.5 & $0.90^*$ & 0.00067 & 0.0678 & 0.0242 & 0.0109 \\ \hline
 4.5 & $0.90^*$ & 0.00160 & 0.0314 & 0.0177 & 0.0044 \\ \hline
 7.5 & 0.04 & 0.00020 & 0.0201 & 0.0053 & 0.0011 \\ \hline
 7.5 & 0.04 & 0.00040 & 0.0195 & 0.0043 & 0.0027 \\ \hline
 7.5 & 0.04 & 0.00079 & 0.0215 & 0.0046 & 0.0039 \\ \hline
 7.5 & 0.10 & 0.00020 & 0.0205 & 0.0071 & 0.0021 \\ \hline
 7.5 & 0.10 & 0.00040 & 0.0195 & 0.0060 & 0.0019 \\ \hline
 7.5 & 0.10 & 0.00079 & 0.0124 & 0.0042 & 0.0013 \\ \hline
 7.5 & 0.10 & 0.00158 & 0.0298 & 0.0074 & 0.0056 \\ \hline
 7.5 & 0.10 & 0.00316 & 0.0490 & 0.0156 & 0.0101 \\ \hline
 7.5 & 0.20 & 0.00020 & 0.0258 & 0.0085 & 0.0019 \\ \hline
 7.5 & 0.20 & 0.00040 & 0.0210 & 0.0065 & 0.0023 \\ \hline
 7.5 & 0.20 & 0.00079 & 0.0174 & 0.0054 & 0.0016 \\ \hline
 7.5 & 0.20 & 0.00158 & 0.0170 & 0.0050 & 0.0021 \\ \hline
 7.5 & 0.20 & 0.00316 & 0.0222 & 0.0065 & 0.0034 \\ \hline
 7.5 & 0.40 & 0.00020 & 0.0354 & 0.0123 & 0.0028 \\ \hline
 7.5 & 0.40 & 0.00040 & 0.0277 & 0.0102 & 0.0014 \\ \hline
 7.5 & 0.40 & 0.00079 & 0.0266 & 0.0092 & 0.0026 \\ \hline
 7.5 & 0.40 & 0.00158 & 0.0155 & 0.0064 & 0.0010 \\ \hline
 7.5 & 0.40 & 0.00316 & 0.0394 & 0.0110 & 0.0048 \\ \hline
 7.5 & 0.40 & 0.00668 & 0.0231 & 0.0083 & 0.0036 \\ \hline
 7.5 & 0.65 & 0.00020 & 0.0764 & 0.0221 & 0.0116 \\ \hline
 7.5 & 0.65 & 0.00040 & 0.0253 & 0.0091 & 0.0026 \\ \hline
 7.5 & 0.65 & 0.00079 & 0.0347 & 0.0120 & 0.0029 \\ \hline
 7.5 & 0.65 & 0.00158 & 0.0196 & 0.0076 & 0.0024 \\ \hline
 7.5 & 0.65 & 0.00316 & 0.0146 & 0.0074 & 0.0025 \\ \hline
 7.5 & 0.65 & 0.00668 & 0.0212 & 0.0082 & 0.0034 \\ \hline
 7.5 & $0.90^*$ & 0.00040 & 0.0638 & 0.0209 & 0.0122 \\ \hline
 7.5 & $0.90^*$ & 0.00079 & 0.0434 & 0.0158 & 0.0063 \\ \hline
 7.5 & $0.90^*$ & 0.00158 & 0.0379 & 0.0148 & 0.0071 \\ \hline
 7.5 & $0.90^*$ & 0.00316 & 0.0365 & 0.0177 & 0.0062 \\ \hline
 7.5 & $0.90^*$ & 0.00668 & 0.0257 & 0.0121 & 0.0077 \\ \hline
 7.5 & $0.90^*$ & 0.01778 & 0.0220 & 0.0116 & 0.0064 \\ \hline
 9.0 & 0.04 & 0.00024 & 0.0279 & 0.0035 & 0.0027 \\ \hline
 9.0 & 0.04 & 0.00042 & 0.0211 & 0.0026 & 0.0020 \\ \hline
 9.0 & 0.04 & 0.00075 & 0.0313 & 0.0040 & 0.0027 \\ \hline
 9.0 & 0.10 & 0.00024 & 0.0252 & 0.0035 & 0.0019 \\ \hline
 9.0 & 0.10 & 0.00042 & 0.0304 & 0.0042 & 0.0033 \\ \hline
 
\end{tabular}}
\hspace{0.01cm}
{\begin{tabular}{|c|c|c|c|c|c|} \hline
 $Q^2$ & $\beta$ & $x$ & $\xpom F_2^{D(3)}$ & $\delta_{\it stat}$ & $\delta_{\it sys}$ \\ \hline
 9.0 & 0.10 & 0.00075 & 0.0261 & 0.0039 & 0.0030 \\ \hline
 9.0 & 0.10 & 0.00133 & 0.0248 & 0.0041 & 0.0029 \\ \hline
 9.0 & 0.10 & 0.00237 & 0.0317 & 0.0057 & 0.0034 \\ \hline
 9.0 & 0.20 & 0.00024 & 0.0260 & 0.0035 & 0.0016 \\ \hline
 9.0 & 0.20 & 0.00042 & 0.0354 & 0.0041 & 0.0022 \\ \hline
 9.0 & 0.20 & 0.00075 & 0.0200 & 0.0026 & 0.0018 \\ \hline
 9.0 & 0.20 & 0.00133 & 0.0257 & 0.0034 & 0.0015 \\ \hline
 9.0 & 0.20 & 0.00237 & 0.0235 & 0.0033 & 0.0028 \\ \hline
 9.0 & 0.20 & 0.00421 & 0.0270 & 0.0046 & 0.0030 \\ \hline
 9.0 & 0.40 & 0.00024 & 0.0456 & 0.0064 & 0.0029 \\ \hline
 9.0 & 0.40 & 0.00042 & 0.0446 & 0.0060 & 0.0028 \\ \hline
 9.0 & 0.40 & 0.00075 & 0.0375 & 0.0053 & 0.0029 \\ \hline
 9.0 & 0.40 & 0.00133 & 0.0226 & 0.0035 & 0.0013 \\ \hline
 9.0 & 0.40 & 0.00237 & 0.0222 & 0.0044 & 0.0014 \\ \hline
 9.0 & 0.40 & 0.00421 & 0.0351 & 0.0066 & 0.0025 \\ \hline
 9.0 & 0.40 & 0.00750 & 0.0247 & 0.0061 & 0.0032 \\ \hline
 9.0 & 0.65 & 0.00024 & 0.0469 & 0.0072 & 0.0059 \\ \hline
 9.0 & 0.65 & 0.00042 & 0.0511 & 0.0065 & 0.0057 \\ \hline
 9.0 & 0.65 & 0.00075 & 0.0378 & 0.0054 & 0.0028 \\ \hline
 9.0 & 0.65 & 0.00133 & 0.0445 & 0.0065 & 0.0051 \\ \hline
 9.0 & 0.65 & 0.00237 & 0.0238 & 0.0047 & 0.0024 \\ \hline
 9.0 & 0.65 & 0.00421 & 0.0222 & 0.0044 & 0.0029 \\ \hline
 9.0 & 0.65 & 0.00750 & 0.0161 & 0.0044 & 0.0030 \\ \hline
 9.0 & $0.90^*$ & 0.00042 & 0.0346 & 0.0061 & 0.0034 \\ \hline
 9.0 & $0.90^*$ & 0.00075 & 0.0332 & 0.0066 & 0.0032 \\ \hline
 9.0 & $0.90^*$ & 0.00133 & 0.0246 & 0.0055 & 0.0030 \\ \hline
 9.0 & $0.90^*$ & 0.00237 & 0.0310 & 0.0068 & 0.0024 \\ \hline
 9.0 & $0.90^*$ & 0.00421 & 0.0277 & 0.0080 & 0.0031 \\ \hline
 9.0 & $0.90^*$ & 0.00750 & 0.0106 & 0.0041 & 0.0012 \\ \hline
12.0 & 0.04 & 0.00024 & 0.0281 & 0.0054 & 0.0050 \\ \hline
12.0 & 0.04 & 0.00042 & 0.0277 & 0.0034 & 0.0023 \\ \hline
12.0 & 0.04 & 0.00075 & 0.0284 & 0.0047 & 0.0028 \\ \hline
12.0 & 0.04 & 0.00133 & 0.0452 & 0.0056 & 0.0056 \\ \hline
12.0 & 0.10 & 0.00024 & 0.0308 & 0.0061 & 0.0028 \\ \hline
12.0 & 0.10 & 0.00042 & 0.0268 & 0.0038 & 0.0022 \\ \hline
12.0 & 0.10 & 0.00075 & 0.0256 & 0.0033 & 0.0017 \\ \hline
12.0 & 0.10 & 0.00133 & 0.0285 & 0.0045 & 0.0030 \\ \hline
12.0 & 0.10 & 0.00237 & 0.0298 & 0.0044 & 0.0033 \\ \hline
12.0 & 0.20 & 0.00024 & 0.0230 & 0.0042 & 0.0015 \\ \hline
12.0 & 0.20 & 0.00042 & 0.0351 & 0.0045 & 0.0030 \\ \hline
12.0 & 0.20 & 0.00075 & 0.0248 & 0.0031 & 0.0018 \\ \hline
12.0 & 0.20 & 0.00133 & 0.0239 & 0.0031 & 0.0015 \\ \hline
12.0 & 0.20 & 0.00237 & 0.0302 & 0.0041 & 0.0022 \\ \hline
12.0 & 0.20 & 0.00421 & 0.0222 & 0.0033 & 0.0022 \\ \hline
12.0 & 0.20 & 0.00750 & 0.0403 & 0.0068 & 0.0068 \\ \hline
12.0 & 0.40 & 0.00024 & 0.0372 & 0.0077 & 0.0033 \\ \hline
12.0 & 0.40 & 0.00042 & 0.0478 & 0.0060 & 0.0039 \\ \hline
12.0 & 0.40 & 0.00075 & 0.0351 & 0.0050 & 0.0026 \\ \hline
12.0 & 0.40 & 0.00133 & 0.0354 & 0.0049 & 0.0025 \\ \hline
12.0 & 0.40 & 0.00237 & 0.0267 & 0.0046 & 0.0026 \\ \hline
12.0 & 0.40 & 0.00421 & 0.0186 & 0.0034 & 0.0018 \\ \hline
12.0 & 0.40 & 0.00750 & 0.0266 & 0.0047 & 0.0047 \\ \hline
12.0 & 0.65 & 0.00024 & 0.0585 & 0.0133 & 0.0100 \\ \hline
12.0 & 0.65 & 0.00042 & 0.0523 & 0.0072 & 0.0044 \\ \hline
12.0 & 0.65 & 0.00075 & 0.0442 & 0.0057 & 0.0045 \\ \hline
12.0 & 0.65 & 0.00133 & 0.0425 & 0.0061 & 0.0043 \\ \hline
 
\end{tabular}}
\end{scriptsize}
\scaption{The quantity $\xpom F_2^{D(3)}$ as a function of $x$, $\beta$ and
  $Q^2$  together with the statistical and systematic errors.  Not
  included in the errors are separate overall normalisation uncertainties of
  $\pm 6.0\%$ for the data with $Q^2<9\,{\rm GeV}^2$ and $\pm4.8\%$ for those
  with $Q^2\ge 9\,{\rm GeV}^2$. The data marked with an asterix correspond to
  the cross section, $\dthreesig$, averaged across the bin in $\beta$. The
  $\beta$ boundaries for these data are 0.5 and 0.8 for the bins at
  $\beta=0.65$, and 0.8 and 1.0 for the bins at $\beta=0.9$.}
\label{tab:results}
\end{table}
\addtocounter{table}{-1}
\begin{table}[h]
\begin{scriptsize}
{\begin{tabular}{|c|c|c|c|c|c|} \hline
 $Q^2$ & $\beta$ & $x$ & $\xpom F_2^{D(3)}$ & $\delta_{\it stat}$ & $\delta_{\it sys}$ \\ \hline
12.0 & 0.65 & 0.00237 & 0.0284 & 0.0045 & 0.0023 \\ \hline
12.0 & 0.65 & 0.00421 & 0.0264 & 0.0044 & 0.0032 \\ \hline
12.0 & 0.65 & 0.00750 & 0.0269 & 0.0054 & 0.0028 \\ \hline
12.0 & $0.90^*$ & 0.00042 & 0.0321 & 0.0064 & 0.0042 \\ \hline
12.0 & $0.90^*$ & 0.00075 & 0.0426 & 0.0076 & 0.0037 \\ \hline
12.0 & $0.90^*$ & 0.00133 & 0.0359 & 0.0075 & 0.0032 \\ \hline
12.0 & $0.90^*$ & 0.00237 & 0.0322 & 0.0064 & 0.0019 \\ \hline
12.0 & $0.90^*$ & 0.00421 & 0.0114 & 0.0033 & 0.0011 \\ \hline
12.0 & $0.90^*$ & 0.00750 & 0.0176 & 0.0050 & 0.0018 \\ \hline
18.0 & 0.04 & 0.00042 & 0.0350 & 0.0039 & 0.0031 \\ \hline
18.0 & 0.04 & 0.00075 & 0.0408 & 0.0039 & 0.0030 \\ \hline
18.0 & 0.04 & 0.00133 & 0.0513 & 0.0054 & 0.0055 \\ \hline
18.0 & 0.10 & 0.00042 & 0.0286 & 0.0041 & 0.0024 \\ \hline
18.0 & 0.10 & 0.00075 & 0.0323 & 0.0038 & 0.0027 \\ \hline
18.0 & 0.10 & 0.00133 & 0.0309 & 0.0038 & 0.0033 \\ \hline
18.0 & 0.10 & 0.00237 & 0.0411 & 0.0047 & 0.0043 \\ \hline
18.0 & 0.10 & 0.00421 & 0.0429 & 0.0066 & 0.0070 \\ \hline
18.0 & 0.20 & 0.00042 & 0.0344 & 0.0042 & 0.0021 \\ \hline
18.0 & 0.20 & 0.00075 & 0.0256 & 0.0030 & 0.0015 \\ \hline
18.0 & 0.20 & 0.00133 & 0.0261 & 0.0030 & 0.0018 \\ \hline
18.0 & 0.20 & 0.00237 & 0.0251 & 0.0028 & 0.0027 \\ \hline
18.0 & 0.20 & 0.00421 & 0.0257 & 0.0029 & 0.0022 \\ \hline
18.0 & 0.20 & 0.00750 & 0.0433 & 0.0065 & 0.0049 \\ \hline
18.0 & 0.40 & 0.00042 & 0.0382 & 0.0055 & 0.0020 \\ \hline
18.0 & 0.40 & 0.00075 & 0.0391 & 0.0051 & 0.0030 \\ \hline
18.0 & 0.40 & 0.00133 & 0.0405 & 0.0050 & 0.0027 \\ \hline
18.0 & 0.40 & 0.00237 & 0.0275 & 0.0037 & 0.0019 \\ \hline
18.0 & 0.40 & 0.00421 & 0.0328 & 0.0045 & 0.0028 \\ \hline
18.0 & 0.40 & 0.00750 & 0.0312 & 0.0042 & 0.0029 \\ \hline
18.0 & 0.40 & 0.01330 & 0.0258 & 0.0049 & 0.0063 \\ \hline
18.0 & 0.65 & 0.00042 & 0.0543 & 0.0074 & 0.0051 \\ \hline
18.0 & 0.65 & 0.00075 & 0.0497 & 0.0064 & 0.0037 \\ \hline
18.0 & 0.65 & 0.00133 & 0.0360 & 0.0046 & 0.0028 \\ \hline
18.0 & 0.65 & 0.00237 & 0.0439 & 0.0056 & 0.0038 \\ \hline
18.0 & 0.65 & 0.00421 & 0.0272 & 0.0038 & 0.0026 \\ \hline
18.0 & 0.65 & 0.00750 & 0.0251 & 0.0038 & 0.0025 \\ \hline
18.0 & 0.65 & 0.01330 & 0.0196 & 0.0044 & 0.0032 \\ \hline
18.0 & $0.90^*$ & 0.00075 & 0.0397 & 0.0066 & 0.0041 \\ \hline
18.0 & $0.90^*$ & 0.00133 & 0.0370 & 0.0065 & 0.0037 \\ \hline
18.0 & $0.90^*$ & 0.00237 & 0.0330 & 0.0057 & 0.0028 \\ \hline
18.0 & $0.90^*$ & 0.00421 & 0.0205 & 0.0044 & 0.0014 \\ \hline
18.0 & $0.90^*$ & 0.00750 & 0.0199 & 0.0056 & 0.0019 \\ \hline
18.0 & $0.90^*$ & 0.01330 & 0.0180 & 0.0062 & 0.0016 \\ \hline
28.0 & 0.04 & 0.00075 & 0.0476 & 0.0050 & 0.0048 \\ \hline
28.0 & 0.04 & 0.00133 & 0.0407 & 0.0051 & 0.0056 \\ \hline
28.0 & 0.10 & 0.00075 & 0.0416 & 0.0060 & 0.0044 \\ \hline
28.0 & 0.10 & 0.00133 & 0.0388 & 0.0056 & 0.0034 \\ \hline
28.0 & 0.10 & 0.00237 & 0.0414 & 0.0058 & 0.0034 \\ \hline
28.0 & 0.10 & 0.00421 & 0.0347 & 0.0064 & 0.0072 \\ \hline
28.0 & 0.20 & 0.00075 & 0.0385 & 0.0052 & 0.0023 \\ \hline
28.0 & 0.20 & 0.00133 & 0.0303 & 0.0039 & 0.0022 \\ \hline
28.0 & 0.20 & 0.00237 & 0.0306 & 0.0039 & 0.0021 \\ \hline
28.0 & 0.20 & 0.00421 & 0.0354 & 0.0045 & 0.0030 \\ \hline
28.0 & 0.20 & 0.00750 & 0.0434 & 0.0096 & 0.0043 \\ \hline
28.0 & 0.40 & 0.00075 & 0.0476 & 0.0074 & 0.0037 \\ \hline
28.0 & 0.40 & 0.00133 & 0.0435 & 0.0063 & 0.0025 \\ \hline
28.0 & 0.40 & 0.00237 & 0.0337 & 0.0050 & 0.0029 \\ \hline
 
\end{tabular}}
\hspace{0.01cm}
{\begin{tabular}{|c|c|c|c|c|c|} \hline
 $Q^2$ & $\beta$ & $x$ & $\xpom F_2^{D(3)}$ & $\delta_{\it stat}$ & $\delta_{\it sys}$ \\ \hline
28.0 & 0.40 & 0.00421 & 0.0401 & 0.0057 & 0.0028 \\ \hline
28.0 & 0.40 & 0.00750 & 0.0331 & 0.0053 & 0.0042 \\ \hline
28.0 & 0.40 & 0.01330 & 0.0470 & 0.0078 & 0.0058 \\ \hline
28.0 & 0.65 & 0.00075 & 0.0539 & 0.0084 & 0.0043 \\ \hline
28.0 & 0.65 & 0.00133 & 0.0451 & 0.0066 & 0.0035 \\ \hline
28.0 & 0.65 & 0.00237 & 0.0335 & 0.0053 & 0.0032 \\ \hline
28.0 & 0.65 & 0.00421 & 0.0236 & 0.0041 & 0.0018 \\ \hline
28.0 & 0.65 & 0.00750 & 0.0341 & 0.0054 & 0.0026 \\ \hline
28.0 & 0.65 & 0.01330 & 0.0217 & 0.0042 & 0.0024 \\ \hline
28.0 & 0.65 & 0.02370 & 0.0223 & 0.0063 & 0.0042 \\ \hline
28.0 & $0.90^*$ & 0.00133 & 0.0321 & 0.0072 & 0.0042 \\ \hline
28.0 & $0.90^*$ & 0.00237 & 0.0249 & 0.0069 & 0.0027 \\ \hline
28.0 & $0.90^*$ & 0.00421 & 0.0260 & 0.0065 & 0.0017 \\ \hline
28.0 & $0.90^*$ & 0.00750 & 0.0249 & 0.0072 & 0.0030 \\ \hline
28.0 & $0.90^*$ & 0.01330 & 0.0140 & 0.0050 & 0.0013 \\ \hline
28.0 & $0.90^*$ & 0.02370 & 0.0129 & 0.0071 & 0.0020 \\ \hline
45.0 & 0.04 & 0.00133 & 0.0585 & 0.0081 & 0.0078 \\ \hline
45.0 & 0.10 & 0.00133 & 0.0384 & 0.0066 & 0.0030 \\ \hline
45.0 & 0.10 & 0.00237 & 0.0328 & 0.0061 & 0.0043 \\ \hline
45.0 & 0.20 & 0.00133 & 0.0268 & 0.0049 & 0.0023 \\ \hline
45.0 & 0.20 & 0.00237 & 0.0406 & 0.0061 & 0.0026 \\ \hline
45.0 & 0.20 & 0.00421 & 0.0302 & 0.0049 & 0.0026 \\ \hline
45.0 & 0.20 & 0.00750 & 0.0310 & 0.0057 & 0.0027 \\ \hline
45.0 & 0.40 & 0.00133 & 0.0347 & 0.0072 & 0.0024 \\ \hline
45.0 & 0.40 & 0.00237 & 0.0387 & 0.0070 & 0.0021 \\ \hline
45.0 & 0.40 & 0.00421 & 0.0215 & 0.0046 & 0.0016 \\ \hline
45.0 & 0.40 & 0.00750 & 0.0273 & 0.0053 & 0.0027 \\ \hline
45.0 & 0.40 & 0.01330 & 0.0328 & 0.0067 & 0.0043 \\ \hline
45.0 & 0.65 & 0.00133 & 0.0418 & 0.0081 & 0.0027 \\ \hline
45.0 & 0.65 & 0.00237 & 0.0262 & 0.0058 & 0.0013 \\ \hline
45.0 & 0.65 & 0.00421 & 0.0294 & 0.0061 & 0.0028 \\ \hline
45.0 & 0.65 & 0.00750 & 0.0240 & 0.0054 & 0.0026 \\ \hline
45.0 & 0.65 & 0.01330 & 0.0199 & 0.0045 & 0.0016 \\ \hline
45.0 & 0.65 & 0.02370 & 0.0220 & 0.0060 & 0.0025 \\ \hline
45.0 & 0.90 & 0.00237 & 0.0291 & 0.0094 & 0.0022 \\ \hline
45.0 & 0.90 & 0.00421 & 0.0204 & 0.0072 & 0.0013 \\ \hline
45.0 & 0.90 & 0.00750 & 0.0088 & 0.0041 & 0.0005 \\ \hline
45.0 & 0.90 & 0.01330 & 0.0201 & 0.0078 & 0.0015 \\ \hline
45.0 & 0.90 & 0.02370 & 0.0031 & 0.0032 & 0.0004 \\ \hline
75.0 & 0.10 & 0.00237 & 0.0453 & 0.0090 & 0.0047 \\ \hline
75.0 & 0.10 & 0.00421 & 0.0610 & 0.0137 & 0.0088 \\ \hline
75.0 & 0.20 & 0.00237 & 0.0341 & 0.0073 & 0.0037 \\ \hline
75.0 & 0.20 & 0.00421 & 0.0407 & 0.0071 & 0.0035 \\ \hline
75.0 & 0.20 & 0.00750 & 0.0312 & 0.0073 & 0.0032 \\ \hline
75.0 & 0.40 & 0.00237 & 0.0444 & 0.0120 & 0.0029 \\ \hline
75.0 & 0.40 & 0.00421 & 0.0405 & 0.0096 & 0.0037 \\ \hline
75.0 & 0.40 & 0.00750 & 0.0236 & 0.0063 & 0.0020 \\ \hline
75.0 & 0.40 & 0.01330 & 0.0318 & 0.0080 & 0.0031 \\ \hline
75.0 & 0.65 & 0.00237 & 0.0224 & 0.0075 & 0.0016 \\ \hline
75.0 & 0.65 & 0.00421 & 0.0243 & 0.0072 & 0.0018 \\ \hline
75.0 & 0.65 & 0.00750 & 0.0314 & 0.0079 & 0.0027 \\ \hline
75.0 & 0.65 & 0.01330 & 0.0212 & 0.0056 & 0.0025 \\ \hline
75.0 & 0.65 & 0.02370 & 0.0252 & 0.0083 & 0.0026 \\ \hline
75.0 & 0.90 & 0.00421 & 0.0234 & 0.0104 & 0.0020 \\ \hline
75.0 & 0.90 & 0.00750 & 0.0109 & 0.0066 & 0.0013 \\ \hline
75.0 & 0.90 & 0.01330 & 0.0145 & 0.0082 & 0.0019 \\ \hline
75.0 & 0.90 & 0.02370 & 0.0093 & 0.0062 & 0.0011 \\ \hline
 
\end{tabular}}
\end{scriptsize}
\scaption{The quantity  $\xpom F_2^{D(3)}$  (cont'd).}
\end{table}
\end{document}